\documentclass{article}%
\usepackage{hyperref}
\usepackage{amsmath}
\usepackage{amsfonts}
\usepackage{amssymb}
\usepackage{graphicx}%
\begin{document}

\title{Field-Dependent BRST-antiBRST Lagrangian Transformations}
\author{\textsc{Pavel Yu. Moshin${}^{a}$\thanks{moshin@rambler.ru \hspace{0.5cm}
${}^{\dagger}$reshet@ispms.tsc.ru}\ \ and Alexander A. Reshetnyak${}%
^{b,c\ddagger}$}\\\textit{${}^{a}$Department of Physics, Tomsk
State University, 634050, Tomsk,
Russia,}\\\textit{${}^{b}$Institute of Strength Physics and
Materials Science,}\\\textit{Siberian Branch of Russian Academy of
Sciences, 634021, Tomsk, Russia,}\\\textit{${}^{c}$Tomsk State
Pedagogical University, 634061, Tomsk, Russia }} \maketitle

\begin{abstract}
We continue our study of finite BRST-antiBRST transformations for
general gauge theories in Lagrangian formalism, initiated in
[arXiv:1405.0790[hep-th] and arXiv:1406.0179[hep-th]], with a
doublet $\lambda_{a}$, $a=1,2$, of anticommuting Grassmann
parameters and prove the correctness of the explicit Jacobian in
the partition function announced in [arXiv:1406.0179[hep-th]],
which corresponds to a change of variables with
functionally-dependent parameters $\lambda_{a}=U_{a}\Lambda$
induced by a finite Bosonic functional $\Lambda(\phi,\pi,\lambda)$
and by the anticommuting generators $U_{a}$ of BRST-antiBRST
transformations in the space of fields $\phi$ and auxiliary
variables $\pi^{a},\lambda$. We obtain a Ward
identity depending on the field-dependent parameters $\lambda_{a}$
and study the problem of gauge dependence, including the case of
Yang--Mills theories. We examine a formulation with BRST-antiBRST
symmetry breaking terms, additively introduced to the quantum
action constructed by the Sp(2)-covariant Lagrangian
rules, obtain the Ward identity and investigate the
gauge-independence of the corresponding generating functional of
Green's functions. A formulation with BRST symmetry breaking terms
is developed. It is argued that the gauge independence of
the above generating functionals is fulfilled in the BRST and
BRST-antiBRST settings. These concepts are applied to the average
effective action in Yang--Mills theories within the functional
renormalization group approach.

\end{abstract}


\section{Introduction}

In our recent works \cite{MRnew,MRnew1,MRnew2}, we have proposed
an extension of BRST-antiBRST transformations to the case of
finite (global and field-dependent) parameters for Yang--Mills and
general gauge theories within the BRST-antiBRST Lagrangian
\cite{BLT1,BLT2,Hull} and generalized Hamiltonian
\cite{BLT1h,BLT2h} quantization schemes; see also \cite{GH1}. The
notion of ``finiteness'' employs the inclusion into finite
transformations of a new term, being quadratic in the parameters
$\mu_{a}$. First of all, this makes it possible to realize the
complete BRST-antiBRST invariance of the integrand in the vacuum
functional. Second, the functionally-dependent parameters
$\lambda_{a}=s_{a}\Lambda$, induced by a Bosonic functional
$\Lambda$, provide an explicit correspondence (due to the
compensation equation for the corresponding Jacobian) between a
choice of $\Lambda$ and a transition from the partition function
of a theory in a certain gauge, determined by a Bosonic gauge
functional $F_{0}$, to the same theory in a different gauge, given
by another gauge Boson $F$. This becomes a key instrument of a
BRST-antiBRST approach that allows one to determine the Gribov
horizon functional \cite{Gribov} -- which is initially given by
the Landau gauge in the Gribov--Zwanziger theory \cite{Zwanziger}
-- by using any other gauge, including the $R_{\xi}$-gauges,
eliminating residual gauge invariance in the deep IR region. In
this respect, it should be noted that we do not consider here, and
have not considered earlier in \cite{MRnew,MRnew1,MRnew2}, the
case of BRST \cite{BRST1,BRST2} and antiBRST transformations with
one and the same anticommuting Grassmann parameter
$\mu=\delta\Lambda$, as suggested in the first paper \cite{JM}
devoted to finite BRST transformations (see Eqs. (2.3a), (2.3b)
therein) in the infinitesimal and finite forms for the global and
field-dependent cases in Yang--Mills theories. In fact, such
considerations are in conflict with the ghost number distribution
used in \cite{MRnew,BLT1,BLT2} for field variables and also
contradict to the standard definition of BRST-antiBRST symmetry
transformations \cite{aBRST1,aBRST2,aBRST3}, which implies the
presence in BRST transformed fields of a Grassmann-odd parameter
$\mu$ and an independent Grassmann-odd parameter for antiBRST
transformed fields, $\bar{\mu}$, with the ghost number opposite to
that of $\mu$.\footnote{According to the authors of \cite{JM},
however, Eqs. (2.3a), (2.3b) apply to the general case, so that
the same parameter $\delta\Lambda$ has to be understood
differently in each of these formulae.} Second, our consideration
is based on special global two-parametric supersymmetries,
realized on equal footing in \cite{BLT1,BLT2,Hull,BLT1h,BLT2h}. At
the same time, finite field-dependent BRST and antiBRST
transformations in Yang--Mills theories and reducible gauge
theories with Abelian gauge groups have been recently examined in
\cite{Upadhyay1,Upadhyay3} using a different scheme at different
stages of quantization, whereas the so-called finite ``mixed
BRST-antiBRST transformations'' -- by the terminology of
\cite{MRnew, MRnew1, MRnew2, BLT1, BLT2, Hull, BLT1h, BLT2h} for
BRST-antiBRST transformations; see (3.7) in \cite{Upadhyay3} -- do
not contain the polynomial term $\Theta_{1}\Theta_{2} \ne 0$, thus
affecting the ``finiteness'' of such finite BRST-antiBRST
transformations. In fact, this eliminates the capability of such
finite BRST-antiBRST transformations to provide symmetry
transformations such that would relate the partition function of a
gauge theory in one gauge to the same theory in another gauge
within perturbation theory. Instead, by making a change of
variables related to such field-dependent transformations in the
vacuum functional (even in Abelian gauge theories), one cannot
preserve the quantum action of a given theory and to obtain this
theory in another gauge, thus making impossible the gauge
independence of the vacuum functional and of the physical
$S$-matrix for a finite change of the gauge
condition.\footnote{Calculations of Jacobians corresponding to
BRST-antiBRST transformations linear in finite field-dependent
parameters for Yang--Mills and more general gauge theories with an
open gauge algebra, as well as transformations with polynomial,
albeit functionally-independent, parameters $\lambda_a$, is an
essential feature of our future research \cite{MRmew4}.}

For completeness, note that finite field-dependent BRST
transformations for general gauge theories in the BV quantization
scheme \cite{BV} have been examined in \cite{BLTfin}, and earlier
in \cite{Reshetnyak}. A construction of finite field-dependent
BRST-antiBRST transformations in the Sp(2)-covariant generalized
Hamiltonian formalism \cite{BLT1h,BLT2h} has been recently
developed \cite{MRnew1} for arbitrary dynamical systems subject to
first-class constraints, along with an explicit construction of
the parameters $\lambda_{a}$ inducing a change of the gauge for
Yang--Mills theories in the class of $R_{\xi}$-like gauges. In the
case of BRST--BFV symmetry \cite{BRST3}, a study of finite
field-dependent BRST--BFV transformations in the generalized
Hamiltonian formalism \cite{BFV,Henneaux1} has been presented in
\cite{BLThf}. In all of these papers, the crucial point has been
the so-called compensation equation, first suggested for finite
BRST transformations in the Yang--Mills theory \cite{LL1} within
the Faddeev--Popov quantization rules \cite{FP}, which establishes
a one-to-one correspondence of field-dependent parameter(s) of
BRST(-antiBRST) transformations with a finite change of the gauge
condition.

In the discussion of \cite{MRnew2}, namely, see Eqs. (6.2), (6.3)
therein, we have announced an explicit Jacobian in the partition
function $Z_F$ which corresponds to a change of variables with
field-dependent (and functionally-dependent) parameters
$\lambda_a$ of finite BRST-antiBRST transformations, on the basis
of which we solve the compensation equation in order to find
$\lambda_a=\lambda_a(\Delta F)$. This provides the gauge
independence of the vacuum functional, $Z_F=Z_{F+\Delta F}$, and
allows one to obtain the Ward identities and study the problem of
gauge dependence for the generating functional of Green's
functions; see Eqs. (6.5)--(6.10) in \cite{MRnew2}. This concept
was used in \cite{MRnew2} to relate (on the basis of
field-dependent BRST-antiBRST transformations) quantum
BRST-antiBRST invariant actions of the  Freedman--Townsend model
(of an antisymmetric non-Abelian tensor field with a reducible
gauge symmetry) in two different  gauges determined by a gauge
Boson quadratic in the fields.

On the other hand, some problems examined in \cite{MRnew}
for  Yang--Mills theories have remained unsolved. In addition, the topical problem
of BRST-antiBRST symmetry breaking in the Sp(2)-covariant
Lagrangian quantization on the basis of finite field-dependent BRST-antiBRST
transformations, as well as the issue of BRST symmetry breaking
in the BV quantization, initiated in \cite{Reshetnyak,llr1,lrr}
on the basis of finite BRST--BV transformations \cite{BLTfin},
have not been considered.

Based on by these reasons, we intend to address the following problems
related to gauge theories in Lagrangian formalism:

\begin{enumerate}
\item calculation of the Jacobian for a change of variables in the partition
function related to \emph{finite field-dependent BRST-antiBRST
transformations} being polynomial in powers of the
Sp(2)-doublet of Grassmann-odd (and functionally-dependent) parameters
$\lambda_{a}= s_{a}\Lambda$, induced by a finite Grassmann-even
functional $\Lambda(\phi,\pi, \lambda)$ and by the Grassmann-odd
generators $s_{a}$ of BRST-antiBRST transformations;

\item derivation of the Ward identities and consideration of gauge dependence
on the basis of the compensation equation for an unknown functional
$\Lambda(\phi,\pi, \lambda)$ generating the Sp(2)-doublet
$\lambda_{a}$, in order to establish a relation of the partition function
$Z_{F}$ (with the quantum action $\mathcal{S}_{F}$ in a certain gauge
determined by a gauge Boson $F$) to another partition function $Z_{F+\Delta F}$
(with the quantum action $\mathcal{S}_{F+\Delta F}$ in a different gauge $F+\Delta F$);

\item application of these considerations to obtain a new form
of the Ward identities and investigation of gauge dependence
in gauge theories with a closed algebra of rank 1, including Yang--Mills theories;

\item introduction of the concept of BRST-antiBRST symmetry breaking
in the Sp(2)-covariant Lagrangian quantization,
derivation of the Ward identities and the study of gauge dependence
on the basis of finite field-dependent BRST-antiBRST transformations;

\item consideration of a new form (as compared to \cite{llr1, lrr})
of BRST symmetry breaking in the BV quantization, derivation of
the Ward identities and the study of gauge dependence
on the basis of finite field-dependent BRST-BV transformations;

\item application of BRST-antiBRST symmetry breaking concept
to the average effective action in Yang--Mills theories
within the functional renormalization group approach.
\end{enumerate}

The work is organized as follows. In Section~\ref{gensetup}, we
bring to mind the general setup of finite BRST-antiBRST Lagrangian
transformations and prove our conjecture \cite{MRnew2} as to the
form of the Jacobian that corresponds to a change of variables
with functionally-dependent parameters,
$\lambda_{a}=s_{a}\Lambda$. In Section~\ref{WIGD}, we examine the
compensation equation, derive a new form of the Ward identities
depending on the functionals $\lambda_{a}$, and investigate the
problem of gauge dependence, including the case of Yang--Mills
theories. We also obtain the Ward identities using the
field-dependent BRST-antiBRST transformations and study the gauge
dependence of the generating functionals of Green's functions for
general gauge theories in Section~\ref{WIGD1} and for Yang--Mills
theories in Section~\ref{WIGD2}. In Section~\ref{BaBsb}, we
introduce the notion of BRST-antiBRST symmetry breaking in the
Sp(2)-covariant Lagrangian quantization, derive the Ward
identities and study gauge dependence on the basis of finite
field-dependent BRST-antiBRST transformations. In
Appendix~\ref{AppA}, we reconsider the concept of BRST symmetry
breaking within the BV quantization scheme, first developed in
\cite{llr1,lrr}. In  Appendix~\ref{AppB}, we introduce, for the
first time in the BRST-antiBRST Lagrangian quantization, an
average effective action and examine it in a way consistent with
the gauge independence of the conventional S-matrix in Yang{Mills
theories using different gauges. We employ the notation of our
previous works \cite{MRnew, MRnew2}. Unless otherwise specified by
an arrow, derivatives with respect to the fields are taken from
the right, and those with respect to the corresponding antifields
are taken from the left. The raising and lowering of
Sp(2)-indices, $s^{a}=\varepsilon ^{ab}s_{b}$,
$s_{a}=\varepsilon_{ab}s^{b}$, is carried out with the help of a
constant antisymmetric tensor $\varepsilon^{ab}$, $\varepsilon
^{ac}\varepsilon_{cb}=\delta_{b}^{a}$, subject to the
normalization condition $\varepsilon^{12}=1$.

\section{Finite Field-Dependent BRST-antiBRST Transformation and its Jacobian}

\label{gensetup}
\renewcommand{\theequation}{\arabic{section}.\arabic{equation}}
\setcounter{equation}{0} In the Discussion of our recent work
\cite{MRnew}, namely, in Eqs. (6.4), (6.5), we have announced the
form of finite BRST-antiBRST transformations $\Delta\Gamma^{p}$
for a general gauge theory in Lagrangian formalism and
subsequently proved \cite{MRnew2} for constant finite
anticommuting
parameters $\lambda_{a}$ that it actually leaves the integrand $\mathcal{I}%
_{\Gamma}^{\left(F\right)  }$ in the partition function $Z_{F}=\int
d\Gamma\ \mathcal{I}_{\Gamma}^{\left(F\right)  }$ invariant to all orders
in powers of $\lambda_{a}$. Namely, the finite BRST-antiBRST transformations%
\footnote{As shown in \cite{MRnew}, the validity of the algebra of
BRST-antiBRST transformations for its generators
$\overleftarrow{s}^a\overleftarrow{s}^b+\overleftarrow{s}^b
\overleftarrow{s}^a=0$, realized in an appropriate space of
variables in Lagrangian \cite{BLT1} and generalized Hamiltonian
formalism \cite{BLT2h} allows one to restore the finite group form
$\Gamma^{\prime}-\Gamma=\Gamma\left(\overleftarrow{s}^a\lambda_a +
(1/4)\overleftarrow{s}^2\lambda^2\right)$, or, identically,
$\Gamma^{\prime}=\Gamma\left(1+\overleftarrow{s}^a\lambda_a +
(1/4)\overleftarrow{s}^2\lambda^2\right)=\Gamma\exp\left(\overleftarrow{s}^a\lambda_a\right)$.
Equivalently, the realization of the generators in terms of odd anticommuting vector
fields,
$\overleftarrow{s}^a\left(\Gamma\right)=\frac{\overleftarrow{\delta}}{\delta \Gamma^p}
(\Gamma^p\overleftarrow{s}^a)$, due to the Frobenius theorem leads to
the same form of finite BRST-antiBRST transformations. Notice that
the finite BRST-antiBRST transformations are, in fact, constructed
from infinitesimal gauge transformations (instead of finite gauge
group transformations) of classical variables in the case of
finite values of gauge parameters.}
\begin{align}
&  \Delta_\lambda\Gamma^{p}=\Gamma^{p}\left(  \overleftarrow{s}{}^{a}\lambda_{a}%
+\frac{1}{4}\overleftarrow{s}{}^{2}\lambda^{2}\right)  \Longrightarrow
\mathcal{I}_{\Gamma+\Delta_\lambda\Gamma}^{\left(F\right)  }=\mathcal{I}_{\Gamma
}^{\left(F\right)  }\ ,\nonumber\\
&  \mathrm{where}\ \ \ \overleftarrow{s}{}^{a}=\frac{\overleftarrow{\delta}%
}{\delta\phi^{A}}\pi^{Aa}+\frac{\overleftarrow{\delta}}{\delta\phi_{Aa}^{\ast
}}S_{,A}\left(  -1\right)  ^{\varepsilon_{A}}-\varepsilon^{ab}\frac
{\overleftarrow{\delta}}{\delta\bar{\phi}_{A}}\phi_{Ab}^{\ast}\left(
-1\right)  ^{\varepsilon_{A}}+\varepsilon^{ab}\frac{\overleftarrow{\delta}%
}{\delta\pi^{Ab}}\lambda^{A}\ ,\ \ \ \overleftarrow{s}{}^{2}=\overleftarrow
{s}{}^{a}\overleftarrow{s}_{a}\ ,\label{Gamma_fin}%
\end{align}
are realized on the coordinates $\Gamma^{p}=(\phi^{A},\phi_{Aa}^{\ast}%
,\bar{\phi}_{A},\pi^{Aa},\lambda^{A})$ of the space of fields $\phi^{A}$,
antifields $(\phi_{Aa}^{\ast},\bar{\phi}_{A})$ and auxiliary variables
$(\pi^{Aa},\lambda^{A})$ used in the Sp(2)-covariant Lagrangian
quantization \cite{BLT1, BLT2}, with the following distribution of Grassmann
parity and ghost number:%
\begin{align}
&  \varepsilon\left(  \Gamma^{p}\right)  =\left(  \varepsilon_{A}%
,\ \varepsilon_{A}+1,\ \varepsilon_{A},\ \varepsilon_{A}+1,\ \varepsilon
_{A}\right)
 \ ,\nonumber\\
&  \mathrm{gh}\left(  \Gamma^{p}\right)  =\left(  \mathrm{gh}(\phi
^{A}),\ (-1)^{a}-\mathrm{gh}(\phi^{A}),\ -\mathrm{gh}(\phi^{A}),\ (-1)^{a+1}%
+\mathrm{gh}(\phi^{A}),\ \mathrm{gh}(\phi^{A})\right)  \ .\label{gencoor}%
\end{align}
In terms of the components, the transformations (\ref{Gamma_fin}) are given by
($S_{,A}\equiv\delta S/\delta\phi^{A}$)%
\begin{align}
\Delta_\lambda\phi^{A} &  =\pi^{Aa}\lambda_{a}+\frac{1}{2}\lambda^{A}\lambda
^{2}%
,\phantom{= \pi^{Aa}\lambda_{a}+ \frac{1}{2}\lambda^{A} \lambda^{2}\quad\quad\ }\Delta_\lambda
\bar{\phi}_{A}\ =\ \varepsilon^{ab}\lambda_{a}\phi_{Ab}^{\ast}+\frac{1}%
{2}S_{,A}\lambda^{2}\ ,\nonumber\\
\Delta_\lambda\pi^{Aa} &  =-\varepsilon^{ab}\lambda^{A}\lambda_{b}%
\ ,\phantom{= \pi^{Aa}\lambda_{a}+ \frac{1}{2}\lambda^{A} \lambda^{2}=\varepsilon^{ab}\lambda^{A}\lambda_{a}}\Delta_\lambda
\lambda^{A}\ =\ 0\ ,\label{exply}\\
\Delta_\lambda\phi_{Aa}^{\ast} &  =\lambda_{a}S_{,A}+\frac{1}{4}\left(  -1\right)
^{\varepsilon_{A}}\left(  \varepsilon_{ab}\frac{\delta^{2}S}{\delta\phi
^{A}\delta\phi^{B}}\pi^{Bb}+\varepsilon_{ab}\frac{\delta S}{\delta\phi^{B}%
}\frac{\delta^{2}S}{\delta\phi^{A}\delta\phi_{Bb}^{\ast}}\left(  -1\right)
^{\varepsilon_{B}}-\phi_{Ba}^{\ast}\frac{\delta^{2}S}{\delta\phi^{A}\delta
\bar{\phi}_{B}}\left(  -1\right)  ^{\varepsilon_{B}}\right)  \lambda
^{2}\ .\nonumber
\end{align}
The operators $\overleftarrow{s}{}^{a}$ in (\ref{Gamma_fin}) are the
generators of (infinitesimal $\lambda_{a}\equiv\mu_{a}$) BRST-antiBRST
transformations \cite{BLT1} for the integrand, $\mathcal{I}_{\Gamma
+\delta\Gamma}^{\left( F\right)  }=\mathcal{I}_{\Gamma}^{\left(F\right)  }$,%
\begin{equation}
\delta\Gamma^{p}=\Gamma^{p}\overleftarrow{s}{}^{a}\mu_{a}=\delta\left(
\phi^{A},\ \phi_{Ab}^{\ast},\ \bar{\phi}_{A},\ \pi^{Ab},\ \lambda^{A}\right)
=\left(  \pi^{Aa},\ \delta_{b}^{a}S_{,A}\left(  -1\right)  ^{\varepsilon_{A}%
},\ \varepsilon^{ab}\phi_{Ab}^{\ast}\left(  -1\right)  ^{\varepsilon_{A}%
+1},\ \varepsilon^{ab}\lambda^{A},\ 0\right)  \mu_{a}\ ,\label{BaBinf}%
\end{equation}
which may be regarded as integrability conditions for the validity of
$\mathcal{I}_{\Gamma+\Delta_\lambda\Gamma}^{\left(F\right)  }=\mathcal{I}%
_{\Gamma}^{\left( F\right)  }$ to all orders in the parameters
$\lambda_{a}$, with the restriction $\overleftarrow{U}{}^{a}$ of the
generators $\overleftarrow{s}{}^{a}$ to the subspace $(\phi^{A},\pi
^{Ab},\lambda^{A})$ being anticommuting and nilpotent:%
\begin{equation}
\overleftarrow{U}{}^{a}=\left.  \overleftarrow{s}{}^{a}\right\vert _{\phi
,\pi,\lambda}=\frac{\overleftarrow{\delta}}{\delta\phi^{A}}\pi^{Aa}%
+\varepsilon^{ab}\frac{\overleftarrow{\delta}}{\delta\pi^{Ab}}\lambda
^{A}\ ,\ \ \ \overleftarrow{U}{}^{a}\overleftarrow{U}{}^{b}+\overleftarrow
{U}{}^{b}\overleftarrow{U}{}^{a}=0\ ,\ \ \ \overleftarrow{U}{}^{a}%
\overleftarrow{U}{}^{b}\overleftarrow{U}{}^{c}=0\ .\label{nilp}%
\end{equation}
This makes it possible to present the generating functional $Z_{F}(J)$ of
Green's functions in \cite{MRnew2, BLT1}, depending on external sources
$J_{A}$, with $\varepsilon(J_{A})=\varepsilon_{A}$, \textrm{gh}$(J_{A}%
)=-\mathrm{gh}(\phi^{A})$, as the path integral%
\begin{align}
&  Z_{F}(J)=\int d\Gamma\;\exp\left\{  \left(  i/\hbar\right)  \left[
\mathcal{S}_{F}\left(  \Gamma\right)  +J_{A}\phi^{A}\right]  \right\}
\ ,\ \ \ \mathcal{S}_{F}=\mathcal{S-}\left(  1/2\right)  F\overleftarrow{U}%
{}^{2}\ ,\ \ \ \overleftarrow{U}{}^{2}\equiv\overleftarrow{U}{}^{a}%
\overleftarrow{U}_{a}\ ,\nonumber\\
&  \mathrm{where}\ \ \ \mathcal{S=}S+\phi_{Aa}^{\ast}\pi^{Aa}+\bar{\phi}%
_{A}\lambda^{A}\ ,\ \ \ S=S\left(  \phi,\phi^{\ast},\bar{\phi}\right)
\ ,\ \ \ F=F\left(  \phi\right)  \ .\label{z(0)}%
\end{align}
Here, $\hbar$ is the Planck constant; the configuration space
$\phi^{A}$, containing the classical fields $A^{i}$ and the
Sp(2)-symmetric ghost-antighost and Nakanishi--Lautrup
fields, depends on the irreducible \cite{BLT1} or reducible
\cite{BLT2} character of a given gauge theory, whereas the
auxiliary fields $(\pi^{Aa},\lambda^{A})$ are required in order
to introduce the gauge by using a gauge-fixing Bosonic functional
$F(\phi)$ with a vanishing ghost number. In its turn, the Bosonic
functional $S(\phi ,\phi^{\ast},\bar{\phi})$ with a vanishing
ghost number is a solution of the
generating equations%
\begin{equation}
\frac{1}{2}(S,S)^{a}+V^{a}S=i\hbar\Delta^{a}S\Longleftrightarrow\left(
\Delta^{a}+\frac{i}{\hbar}V^{a}\right)  \exp\left(  \frac{i}{\hbar}S\right)
=0\ ,\label{3.3}%
\end{equation}
with a boundary condition for vanishing antifields $\phi_{Aa}^{\ast}$,
$\bar{\phi}_{A}$ given by the classical action $S_{0}(A)$. In (\ref{3.3}), the
extended antibracket $(G,H)^{a}$ for arbitrary functionals $G$, $H$ and the
operators $\Delta^{a}$, $V^{a}$ are given by%
\begin{equation}
(G,H)^{a}=G\left(  \frac{\overleftarrow{\delta}}{\delta\phi^{A}}%
\frac{\overrightarrow{\delta}}{\delta\phi_{Aa}^{\ast}}-\frac{\overleftarrow
{\delta}}{\delta\phi_{Aa}^{\ast}}\frac{\overrightarrow{\delta}}{\delta\phi
^{A}}\right)  H\ ,\ \ \ \Delta^{a}=(-1)^{\varepsilon_{A}}\frac{\overrightarrow
{\delta}}{\delta\phi^{A}}\frac{\overrightarrow{\delta}}{\delta\phi_{Aa}^{\ast
}}\ ,\ \ \ V^{a}=\varepsilon^{ab}\phi_{Ab}^{\ast}\frac{\overrightarrow{\delta
}}{\delta\bar{\phi}_{A}}\ .\label{abrack}%
\end{equation}
The invariance \cite{MRnew2} of the integrand $\mathcal{I}_{\Gamma}^{\left(
F\right)  }=d\Gamma\exp\left[  \left(  i/\hbar\right)  \mathcal{S}_{F}\left(
\Gamma\right)  \right]  $ in (\ref{z(0)}) with vanishing sources $J_{A}=0$
under the global finite BRST-antiBRST transformations (\ref{Gamma_fin}) can be
established by using the generating equations (\ref{3.3}), while taking into
account the nilpotency $\overleftarrow{U}{}^{a}\overleftarrow{U}{}%
^{b}\overleftarrow{U}{}^{c}=0$ of the operators $\overleftarrow{U}{}^{a}$ in
(\ref{z(0)}) and using the explicit change \cite{MRnew2}%
\begin{equation}
\Delta_\lambda G=G\left(  \overleftarrow{s}{}^{a}\lambda_{a}+\frac{1}{4}%
\overleftarrow{s}{}^{2}\lambda^{2}\right)  \label{DeltaF}%
\end{equation}
under the transformations (\ref{Gamma_fin}) of an arbitrary functional
$G\left(  \Gamma\right)  $ expandable as a power series in $\Gamma^{p}$,%
\begin{equation}
G\left(  \Gamma+\Delta_\lambda\Gamma\right)  =G\left(  \Gamma\right)  +G_{,p}\left(
\Gamma\right)  \Delta_\lambda\Gamma^{p}+\left(  1/2\right)  G_{,pq}\left(
\Gamma\right)  \Delta_\lambda\Gamma^{q}\Delta_\lambda\Gamma^{p}=G\left(  \Gamma\right)
+\Delta_\lambda G\left(  \Gamma\right)  \ ,\ \ \ \mathrm{with\ \ \ }G_{,p}\equiv
G\frac{\overleftarrow{\delta}}{\delta\Gamma^{p}}\ ,\label{Gexpan}%
\end{equation}
with allowance for the explicit form \cite{MRnew2} of the Jacobian
$\mathrm{Sdet}\left\Vert \left(  \Gamma^{p}+\Delta_\lambda\Gamma^{p}\right)
\frac{\overleftarrow{\delta}}{\delta\Gamma^{q}}\right\Vert \equiv\exp\left(
\Im\right)  $%
\begin{equation}
\exp\left(  \Im\right)  =\exp\left[  -\left(  \Delta^{a}S\right)  \lambda
_{a}-\frac{1}{4}\left(  \Delta^{a}S\right)  \overleftarrow{s}_{a}\lambda
^{2}\right]  \label{ansjacob}%
\end{equation}
corresponding to the transformation of the integration measure $d\Gamma$ with
respect to the change of variables $\Gamma\rightarrow\check{\Gamma}%
=\Gamma+\Delta_\lambda\Gamma$, namely,%
\begin{align}
&  d\check{\Gamma}=\mathrm{Sdet}\left\Vert \left(  \Gamma^{p}+\Delta_\lambda\Gamma
^{p}\right)  \frac{\overleftarrow{\delta}}{\delta\Gamma^{q}}\right\Vert
=d\Gamma\ \exp\left[  \mathrm{Str\ln}\left(  \mathbb{I+}M\right)  \right]
\equiv d\Gamma\ \exp\left(  \Im\right)  \ ,\nonumber\\
&  \ \mathrm{where}\ \ \ \Im=\mathrm{Str\ln}\left(  \mathbb{I+}M\right)
=-\sum_{n=1}^{\infty}\frac{\left(  -1\right)  ^{n}}{n}\mathrm{Str\ }%
M^{n}\ .\label{superJ}%
\end{align}
In \cite{MRnew2}, we have announced that the Jacobian $\exp\left(
\Im\right) $ and the related integration measure $d\check{\Gamma}$
corresponding to a finite change of variables
$\Gamma\rightarrow\check{\Gamma}$ with the choice of
field-dependent parameters $\lambda_{a}=s_{a}\Lambda$ for
$\Lambda=\Lambda\left(  \phi,\pi,\lambda\right)  $, inspired by
the infinitesimal field-dependent BRST-antiBRST transformations of
\cite{MRnew,BLT1}, should take the form%
\begin{align}
&  \exp\left(  \Im\right)  =\exp\left[  -\left(  \Delta^{a}S\right)
\lambda_{a}-\frac{1}{4}\left(  \Delta^{a}S\right)  \overleftarrow{s}%
_{a}\lambda^{2}\right]  \exp\left[  \ln\left(  1+f\right)  ^{-2}\right]
\ ,\ \ \mathrm{with}\ \ \,f=-\frac{1}{2}\Lambda\overleftarrow{s}{}%
^{2},\label{superJaux}\\
&  d\check{\Gamma}=d\Gamma\ \exp\left[  \frac{i}{\hbar}\left(  -i\hbar
\Im\right)  \right]  =d\Gamma\ \exp\left\{  \frac{i}{\hbar}\left[  {i}{\hbar
}\left(  \Delta^{a}S\right)  \lambda_{a}+\frac{i\hbar}{4}\left(  \Delta
^{a}S\right)  \overleftarrow{s}_{a}\lambda^{2}+i\hbar\,\mathrm{\ln}\left(
1-\frac{1}{2}\Lambda\overleftarrow{s}{}^{2}\right)  ^{2}\right]  \right\}
\ ,\label{superJ1}%
\end{align}
where $\Lambda\left(  \phi,\pi,\lambda\right)  $ is a certain
Bosonic potential with a vanishing ghost number. Therefore, in
order to prove the above statement (\ref{superJaux}),
(\ref{superJ1}), let us examine the general case of a finite
BRST-antiBRST transformation (\ref{Gamma_fin}) parameterized by
$\lambda_{a}\left(  \Gamma\right)  $ and consider the even matrix
$M$ in
(\ref{superJ}) with the elements $M_{q}^{p}$, $\varepsilon\left(  M_{q}%
^{p}\right)  =\varepsilon_{p}+\varepsilon_{q}$,%
\begin{align}
&  M_{q}^{p}=\Delta\Gamma^{p}\frac{\overleftarrow{\delta}}{\delta\Gamma^{q}%
}=U_{q}^{p}+V_{q}^{p}+W_{q}^{p}\,,\,\,\,\mathrm{with}\,\,\,V_{q}^{p}=\left(
V_{1}\right)  _{q}^{p}+\left(  V_{2}\right)  _{q}^{p}\ ,\label{MABext}\\
&  \mathrm{for}\,\,\,U_{q}^{p}=X^{pa}\lambda_{a,q}\ ,\ \ \ \left(
V_{1}\right)  _{q}^{p}=\lambda_{a}X_{,q}^{pa}\left(  -1\right)  ^{\varepsilon
_{p}+1}\ ,\ \ \ \left(  V_{2}\right)  _{q}^{p}=\lambda_{a}Y^{p}\lambda
_{,q}^{a}\left(  -1\right)  ^{\varepsilon_{p}+1}\ ,\ \ \ W_{q}^{p}=-\frac
{1}{2}\lambda^{2}Y_{,q}^{p}\ ,\label{PRQABext}%
\end{align}
taking account of the notation \cite{MRnew2}%
\begin{equation}
{X}^{pa}\equiv\Gamma^{p}\overleftarrow{s}{}^{a}\ \ \ \mathrm{and}\ \ \ {Y}%
^{p}\equiv\left(  1/2\right)  {X}_{,q}^{pa}{X}^{qb}\varepsilon_{ba}=-\left(
1/2\right)  \Gamma^{p}\overleftarrow{s}{}^{2}\ .\label{xy2}%
\end{equation}
First of all, let us establish a useful relation between the matrices $V_{1}$
and $W$ in (\ref{MABext}). To do so, we use the generating equations
(\ref{3.3}) and represent the condition of invariance of the integrand
$\mathcal{I}_{\Gamma}^{\left(  _{F}\right)  }$ in (\ref{z(0)}) under the
BRST-antiBRST transformations $\delta\Gamma^{p}=\Gamma^{p}\overleftarrow{s}%
{}^{a}\mu_{a}=X^{pa}\mu_{a}$ given by (\ref{BaBinf}) in the form%
\begin{equation}
\mathcal{S}_{F,p}X^{pa}=i\hbar X_{,p}^{pa}\ ,\ \ \mathrm{where}\ \
\ X_{,p}^{pa}=-\Delta^{a}S\ .\label{eqinv}%
\end{equation}
Let us now write identically:%
\begin{align*}
\mathrm{Str}\left(  V_{1}\right)  +\mathrm{Str}\left(  W\right)  -\frac{1}%
{2}\mathrm{Str}\left(  V_{1}^{2}\right)    & =\left[  \left(  V_{1}\right)
_{p}^{p}+W_{p}^{p}-\frac{1}{2}\left(  V_{1}\right)  _{q}^{p}\left(
V_{1}\right)  _{p}^{q}\right]  \left(  -1\right)  ^{\varepsilon_{p}}\\
& =X_{,p}^{pa}\lambda_{a}-\frac{1}{2}\left(  -1\right)  ^{\varepsilon_{p}%
}\left(  Y_{,p}^{p}-\frac{1}{2}X_{,q}^{pa}X_{,p}^{qb}\varepsilon_{ba}\right)
\lambda^{2}\ .
\end{align*}
Considering%
\begin{align}
Y_{,p}^{p}-\frac{1}{2}X_{,q}^{pa}X_{,p}^{qb}\varepsilon_{ba} &  =\frac{1}%
{2}\varepsilon_{ba}\left(  X_{,qp}^{pa}X^{qb}\left(  -1\right)  ^{\varepsilon
_{p}\left(  \varepsilon_{q}+1\right)  }+X_{,q}^{pa}X_{,p}^{qb}\right)
-\frac{1}{2}\varepsilon_{ba}X_{,q}^{pa}X_{,p}^{qb}\nonumber\\
&  =\frac{1}{2}\varepsilon_{ba}\left(  X_{,qp}^{pa}X^{qb}\left(  -1\right)
^{\varepsilon_{p}\left(  \varepsilon_{q}+1\right)  }+X_{,q}^{pa}X_{,p}%
^{qb}-X_{,q}^{pa}X_{,p}^{qb}\right)  =\frac{1}{2}\varepsilon_{ba}X_{,pq}%
^{pa}X^{qb}\left(  -1\right)  ^{\varepsilon_{p}}\ ,\label{consstrmpq}%
\end{align}
we arrive at%
\begin{equation}
\mathrm{Str}\left(  V_{1}\right)  +\mathrm{Str}\left(  W\right)  -\frac{1}%
{2}\mathrm{Str}\left(  V_{1}^{2}\right)  =X_{,p}^{pa}\lambda_{a}+\frac{1}%
{4}\varepsilon_{ab}X_{,pq}^{pa}X^{qb}\lambda^{2}\ ,\label{lnjacob}%
\end{equation}
where (\ref{eqinv}) implies ($G\overleftarrow{s}{}^{a}\equiv s^a G$)%
\begin{equation}
X_{,p}^{pa}=-\Delta^{a}S\ ,\ \ \ X_{,pq}^{pa}X^{qb}=-\left(  \Delta
^{a}S\right)  _{,p}X^{pb}=-s^{b}\left(  \Delta^{a}S\right)
\ ,\ \ \mathrm{with}\ \ G_{,p}X^{pa}=G_{,p}\left(  s^{a}\Gamma^{p}\right)
=s^{a}G\ .\label{diffcons}%
\end{equation}
Therefore,
\begin{equation}
\mathrm{Str}\left(  V_{1}\right)  +\mathrm{Str}\left(  W\right)  -\frac{1}%
{2}\mathrm{Str}\left(  V_{1}^{2}\right)  =-\left(  \Delta^{a}S\right)
\lambda_{a}-\frac{1}{4}\left(  s_{a}\Delta^{a}S\right)  \lambda^{2}%
\ .\label{zero}%
\end{equation}
Taking account of the relation between the matrices $V_{1}$ and
$W$ in (\ref{MABext}), established for arbitrary
$\lambda_a(\Gamma)$, we now proceed to the case of field-dependent
parameters $\lambda_{a}=\Lambda\overleftarrow{s}_{a}$ in
(\ref{Gamma_fin}), determined by a Bosonic potential
$\Lambda\left( \phi,\pi,\lambda\right)  $, which implies
$\lambda_{a}=\Lambda\overleftarrow{U}_{a}$, in view of
(\ref{nilp}). To do so, using the property $\mathrm{Str}\left(
AB\right)  $ = $\mathrm{Str}\left(  BA\right)$ of arbitrary even
matrices $A$, $B$ and the fact that the occurrence of
$W\sim\lambda^{2}$ in $\mathrm{Str}\left(
M^{n}\right)  $ more than once yields zero, $\lambda^{4}\equiv0$, we have%
\begin{equation}
\mathrm{Str}\left(  M^{n}\right)  =\mathrm{Str}\left(  U+V+W\right)  ^{n}%
=\sum_{k=0}^{1}C_{n}^{k}\mathrm{Str}\left[  \left(  U+V\right)  ^{n-k}%
W^{k}\right]  \ ,\ \ \ C_{n}^{k}=\frac{n!}{k!\left(  n-k\right)
!}\ .\label{Strgen}%
\end{equation}
Furthermore,%
\begin{equation}
\mathrm{Str}\left(  U+V+W\right)  ^{n}=\mathrm{Str}\left(  U+V\right)
^{n}+n\mathrm{Str}\left[  \left(  U+V\right)  ^{n-1}W\right]  =\mathrm{Str}%
\left(  U+V\right)  ^{n}+n\mathrm{Str}\left(  U^{n-1}W\right)  \ ,\label{R}%
\end{equation}
since any occurrence of $W\sim\lambda^{2}$ and $V\sim\lambda_{a}$
simultaneously entering $\mathrm{Str}\left(  M\right)  ^{n}$\ yields zero,
owing to $\lambda_{a}\lambda^{2}=0$, as a consequence of which $W$ can only be
coupled with $U^{n-1}$. Having established (\ref{R}), let us examine
$\mathrm{Str}\left(  U^{n-1}W\right)  $, namely,%
\begin{equation}
\mathrm{Str}\left(  U^{n-1}W\right)  =\left\{
\begin{array}
[c]{ll}%
\mathrm{Str}\left(  W\right)  \ , & n=1\ ,\\
0\ , & n>1\ .
\end{array}
\right.  \label{StrPR}%
\end{equation}
Indeed, due to the contraction property $U^{2}=f\cdot U\Longrightarrow
U^{l}=f^{l-1}\cdot U$, where $f$ is a Bosonic parameter (for details, see
(\ref{contracprop}) below), we have%
\begin{align}
&  \mathrm{Str}\left(  U^{n-1}W\right)  =f^{n-2}\mathrm{Str}\left(  UW\right)
\ ,\ \ n>1\ ,\label{StrPR1}\\
&  \mathrm{Str}\left(  UW\right)  =\mathrm{Str}\left(  WU\right)  =\left(
WU\right)  _{p}^{p}\left(  -1\right)  ^{\varepsilon_{p}}=W_{q}^{p}U_{p}%
^{q}\left(  -1\right)  ^{\varepsilon_{p}}=-\frac{1}{2}\lambda^{2}\left(
Y_{,q}^{p}X^{qa}\right)  \lambda_{a,p}\left(  -1\right)  ^{\varepsilon_{p}%
}=0\ ,\label{StrPR2}%
\end{align}
since, taking account of the restricted dependence of
$\lambda_{a}(\phi ,\pi,\lambda)$ on $\Gamma^{p}$, the nilpotency
(\ref{nilp}) of the operators $\overleftarrow {U}{}^{a}$, being
the restriction of $\overleftarrow{s}{}^{a}$ to the
subspace $(\phi^{A},\pi^{Aa},\lambda^{A})$, and using the
definitions (\ref{xy2}), we
have%
\begin{align*}
\left(  Y_{,q}^{p}X^{qa}\right)  \lambda_{a,p}\left(  -1\right)
^{\varepsilon_{p}} &
=Y^{p}\overleftarrow{s}{}^{a}\lambda_{a,p}\left(
-1\right)  ^{\varepsilon_{p}}\\
&  =-\frac{1}{2}\left(  \Gamma^{p}\overleftarrow{s}{}^{2}\overleftarrow{s}%
{}^{a}\right)  \lambda_{a,p}\left(  -1\right)  ^{\varepsilon_{p}}=-\frac{1}%
{2}\left(
\Gamma^{p}\overleftarrow{U}{}^{2}\overleftarrow{U}{}^{a}\right)
\lambda_{a,p}\left(  -1\right)  ^{\varepsilon_{p}}=0,
\end{align*}
which implies%
\begin{equation}
\mathrm{Str}\left(  M^{n}\right)  =\mathrm{Str}\left(  U+V\right)
^{n}+n\mathrm{Str}\left(  U^{n-1}W\right)  =\left\{
\begin{array}
[c]{ll}%
\mathrm{Str}\left(  U+V\right)  +\mathrm{Str}\left(  W\right)  \ , & n=1\ ,\\
\mathrm{Str}\left(  U+V\right)  ^{n}\ , & n>1\ ,
\end{array}
\right.  \label{P+Q}%
\end{equation}
so that $W$ drops out of $\mathrm{Str}\left(  M^{n}\right)  $, $n>1,$ and
enters the Jacobian only as $\mathrm{Str}\left(  W\right)  $. Next, as we
examine the contribution $\mathrm{Str}\left(  U+V\right)  ^{n}$ in
(\ref{P+Q}), we notice that an occurrence of $V\sim\lambda_{a}$ more then
twice yields zero, $\lambda_{a}\lambda_{b}\lambda_{c}\equiv0$. Direct
calculations for $n=2,3$ lead to%
\begin{equation}
\mathrm{Str}\left(  U+V\right)  ^{n}=\sum_{k=0}^{n}C_{n}^{k}\mathrm{Str}%
\left(  U^{n-k}V^{k}\right)  =\mathrm{Str}\left(  U^{n}+nU^{n-1}V+C_{n}%
^{2}P^{n-2}V^{2}\right)  \ .\label{StrP+Q}%
\end{equation}
Using considerations identical with those presented in Appendix
B.2 of \cite{MRnew}, we can prove by induction, taking account of
the fact $V\sim\lambda_{a}$ and the contraction property
$U^{2}=f\cdot U$, established in
(\ref{contracprop}), that for any $n\geq4\ $we have%
\begin{equation}
\mathrm{Str}\left(  U+V\right)  ^{n}=\mathrm{Str}\left(  U^{n}+nU^{n-1}%
V+nU^{n-2}V^{2}+K_{n}U^{n-3}VUV\right)  \ ,\label{Str(P+Q)^n}%
\end{equation}
where the coefficients $K_{n}$ are given by%
\begin{equation}
K_{n}=C_{n}^{2}-n\ ,\ \ \ C_{n}^{2}=n\left(  n-1\right)  /2\ \Longrightarrow
\ K_{n}=n\left(  n-3\right)  /2\ ,\label{Kn}%
\end{equation}
which implies%
\begin{equation}
\frac{C_{n}^{2}}{n}-\frac{K_{n}}{n}=1\ ,\ \ \ \frac{C_{n}^{2}}{n}%
-\frac{K_{n+1}}{n+1}=\frac{1}{2}\ .\label{C,K}%
\end{equation}
According to the previous considerations,%
\begin{align}
&  \mathrm{Str}\left(  M^{n}\right)  =\sum_{k=0}^{1}C_{n}^{k}\mathrm{Str}%
\left(  U^{n-k}V^{k}\right)  +D_{n}\ ,\ \ \ n\geq1\ ,\label{M^n}\\
&  \mathrm{for}\,\,\,D_{n}=\left\{
\begin{array}
[c]{ll}%
\mathrm{Str}\left(  W\right)  \ , & n=1\ ,\\
C_{n}^{2}\mathrm{Str}\left(  U^{n-2}V^{2}\right)  \ , & n=2,3\ ,\\
\left(  C_{n}^{2}-K_{n}\right)  \mathrm{Str}\left(  U^{n-2}V^{2}\right)
+K_{n}\mathrm{Str}\left(  U^{n-3}VUV\right)  \ , & n>3\ ,
\end{array}
\right.  \label{D_n}%
\end{align}
we have%
\begin{equation}
\mathrm{Str}\left(  M^{n}\right)  =\left\{
\begin{array}
[c]{ll}%
\mathrm{Str}\left(  U\right)  +\mathrm{Str}\left(  V\right)  +\mathrm{Str}%
\left(  W\right)  \ , & n=1\ ,\\
\mathrm{Str}\left(  U^{n}\right)  +C_{n}^{1}\mathrm{Str}\left(  U^{n-1}%
V\right)  +C_{n}^{2}\mathrm{Str}\left(  U^{n-2}V^{2}\right)  \ , & n=2,3\ ,\\
\mathrm{Str}\left(  U^{n}\right)  +C_{n}^{1}\mathrm{Str}\left(  U^{n-1}%
V\right)  +\left(  C_{n}^{2}-K_{n}\right)  \mathrm{Str}\left(  U^{n-2}%
V^{2}\right)  +K_{n}\mathrm{Str}\left(  U^{n-3}VUV\right)  \ , & n>3\ .
\end{array}
\right.  \label{str(M^n)}%
\end{equation}
First of all, the calculation of the Jacobian is based on the previously
established relation (\ref{zero}) between the matrices $V_{1}$, $W$, and
therefore we will take account of the related combination%
\[
\mathrm{Str}\left(  V_{1}\right)  +\mathrm{Str}\left(  W\right)  -\frac{1}%
{2}\mathrm{Str}\left(  V_{1}^{2}\right)  \ .
\]
Besides, recalling that $\lambda_{a}=\Lambda\overleftarrow{U}_{a}$, we can
deduce the additional property%
\begin{equation}
U^{2}=f\cdot U\ ,\ \ \ f=-\frac{1}{2}\mathrm{Str}\left(  U\right)
\ ,\label{contracprop}%
\end{equation}
where the quantity $f$ is given by%
\begin{equation}
\lambda_{b,p}X^{pa}=\lambda_{b}\overleftarrow{U}{}^{a}=\delta_{b}%
^{a}f\Longrightarrow f=\frac{1}{2}\lambda_{a}\overleftarrow{U}{}^{a}=-\frac
{1}{2}\Lambda\overleftarrow{U}{}^{2}\ .\label{f}%
\end{equation}
Indeed,%
\begin{align}
&  \left(  U^{2}\right)  _{q}^{p}=\left(  U\right)  _{r}^{p}\left(  U\right)
_{q}^{r}=X^{pa}\left(  \lambda_{a,r}X^{rb}\right)  \lambda_{b,q}=f\cdot
\delta_{a}^{b}X^{pa}\lambda_{b,q}=f\cdot\left(  U\right)  _{q}^{p}%
\ ,\nonumber\\
&  \lambda_{a,q}X^{qb}=\lambda_{a}\overleftarrow{s}{}^{b}=\lambda
_{a}\overleftarrow{U}{}^{b}=\Lambda\overleftarrow{U}_{a}\overleftarrow{U}%
{}^{b}=\delta_{a}^{b}f\ ,\ \ \ f=-F_{,A}\lambda^{A}-\left(  1/2\right)
\varepsilon_{ab}\pi^{Aa}F_{,AB}\pi^{Bb}\ ,\nonumber\\
&  f=\frac{1}{2}\lambda_{a,p}X^{pa}=-\frac{1}{2}U_{p}^{p}\left(  -1\right)
^{\varepsilon_{p}}=-\frac{1}{2}\mathrm{Str}\left(  U\right)  \ .\label{fP2}%
\end{align}
As a consequence,%
\begin{align}
&  \mathrm{Str\ }\left(  UV\right)  =\mathrm{Str\ }\left(  VU\right)  =\left(
1+f\right)  \mathrm{Str}\left(V_{2}\right)\ ,\label{prop1}\\
&  \mathrm{Str}\left(  UV^{2}\right)  =\mathrm{Str}\left(
VUV\right) =\left(  1+f\right)  \mathrm{Str}\left[\left(
V_{1}+V_{2}\right)V_{2}\right]
\ ,\label{prop2}\\
&  \mathrm{Str}\left(  UVUV\right)  =\mathrm{Str}\left(  VUVU\right)  =\left(
1+f\right)  ^{2}\mathrm{Str}\left(V^2_{2}\right)  \ .\label{prop3}%
\end{align}
Indeed, due to the relations (\ref{xy2}), (\ref{f}) and their consequences ($G\overleftarrow{U}{}^{a}\equiv U^a G$) %
\begin{equation}
{U}^{a}U^{b}\Gamma^{\underline{p}}=-\varepsilon^{ab}Y^{\underline{p}%
}\ ,\ \ \ U^{a}\lambda^{b}=-\varepsilon^{ab}f\ ,\label{cons}%
\end{equation}
with account taken of the notation%
\begin{equation}
\Gamma^{\underline{p}}=\left(  \phi^{A},\pi^{Aa},\lambda^{A}\right)
\ ,\label{not}%
\end{equation}
we have%
\begin{align}
\mathrm{Str\ }VU &  =\left(  VU\right)  _{p}^{p}\left(  -1\right)
^{\varepsilon_{p}}=V_{r}^{p}U_{p}^{r}\left(  -1\right)  ^{\varepsilon_{p}%
}=-\lambda_{a}\left(  X_{,r}^{pa}+Y^{p}\lambda_{,r}^{a}\right)  X^{rb}%
\lambda_{b,p}\nonumber\\
&  =-\lambda_{a}\left[  \left(  s^{b}s^{a}\Gamma^{\underline{p}}\right)
+Y^{\underline{p}}\left(  s^{b}\lambda^{a}\right)  \right]  \lambda_{b,\underline{p}%
}=-\lambda_{a}\left[  \left(
U^{b}U^{a}\Gamma^{\underline{p}}\right) +Y^{\underline{p}}\left(
U^{b}\lambda^{a}\right)  \right]  \lambda
_{b,\underline{p}}\nonumber\\
&  =-\lambda_{a}\left(  \varepsilon^{ab}Y^{p}+\varepsilon
^{ab}Y^{p}f\right)  \lambda_{b,p}=-\left(  1+f\right)  \lambda_{a}%
Y^{p}f\lambda_{,p}^{a}=\left(  1+f\right)  \mathrm{Str\ }V_{2}\ ,\label{fQ2}%
\end{align}
which proves (\ref{prop1}). In a similar way, using (\ref{cons}),
(\ref{not}) and the property of nilpotency
$\lambda_{a}\lambda_{b}\lambda_{c}=0$, it is straightforward to
verify the remaining properties (\ref{prop2}), (\ref{prop3}) of
the matrices $U$, $V$. As a consequence of (\ref{contracprop}) and
(\ref{prop1})--(\ref{prop3}), we have%
\begin{equation}%
\begin{array}
[c]{lc}%
\mathrm{Str}\left(  U^{n}\right)  =f^{n-1}\mathrm{Str}\left(  U\right)
=-2f^{n}\ , & n\geq1\ ,\\
\mathrm{Str}\left(  U^{n-1}V\right)  =\left\{
\begin{array}
[c]{l}%
\mathrm{Str}\left(  V\right)=\mathrm{Str}\left(V_{1}\right)+\mathrm{Str}\left(V_{2}\right)\ ,\\
f^{n-2}\mathrm{Str}\left(  UV\right) =f^{n-2}\left(  1+f\right)
\mathrm{Str}\ V_{2}\ ,
\end{array}
\right.   &
\begin{array}
[c]{l}%
n=1\ ,\\
n>1\ ,
\end{array}
\\
\mathrm{Str}\left(  U^{n-2}V^{2}\right)  =\left\{
\begin{array}
[c]{l}%
\mathrm{Str}\left(  V^{2}\right)
=\mathrm{Str}\left(V_{1}^{2}\right) +2\mathrm{Str}\left(V_{1}V_{2}\right)
+\mathrm{Str}\left(V_{2}^{2}\right)\ ,\\
f^{n-3}\mathrm{Str}\left(  UV^{2}\right) = f^{n-3}\left(
1+f\right) \mathrm{Str}\left[  \left(  V_{1}+V_{2}\right)
V_{2}\right]  \ ,
\end{array}
\right.   &
\begin{array}
[c]{l}%
n=2\ ,\\
n>2\ ,
\end{array}
\\
\mathrm{Str}\left(  U^{n-3}VUV\right)  =f^{n-4}\mathrm{Str}\left(
UVUV\right)  =f^{n-4}\left(  1+f\right)  ^{2}\mathrm{Str}\ V_{2}^{2}\ , &
n>3\ .
\end{array}
\label{strPnQk}%
\end{equation}
We further notice that $\mathrm{Str}\left(  V_{1}V_{2}\right)  \not \equiv 0$.
Indeed, in view of the definitions (\ref{xy2}) and the nilpotency property
$U^{a}U^{b}U^{c}=0$, we have%
\begin{align}
\left(  V_{1}V_{2}\right)  _{p}^{p}\left(  -1\right)  ^{\varepsilon_{p}} &
=\lambda_{a}X_{,q}^{pa}Y^{q}(\lambda^{2})_{,p}=\frac{1}{2}\lambda_{a}\left(
X_{,q}^{pa}X_{,r}^{qb}\right)  X^{rd}\varepsilon_{db}(\lambda^{2}%
)_{,p}\nonumber\\
&  =\frac{1}{2}\lambda_{a}\left[  (X_{,q}^{\underline{p}a}X^{qb})_{,r}%
-(X_{,q}^{\underline{p}a})_{,r}X^{qb}\left(  -1\right)  ^{\varepsilon
_{r}\left(  \varepsilon_{q}+1\right)  }\right]  X^{rd}\varepsilon_{db}%
(\lambda^{2})_{,\underline{p}}\nonumber\\
&  =\frac{1}{2}\lambda_{a}\left[  (U^{d}U^{b}U^{a}\Gamma^{p})-X_{,qr}%
^{pa}X^{qb}X^{rd}\left(  -1\right)  ^{\varepsilon_{r}\left(  \varepsilon
_{q}+1\right)  }\right]  \varepsilon_{db}(\lambda^{2})_{,p}\nonumber\\
&  =-\frac{1}{2}\varepsilon_{bd}X^{qb}X_{,rq}^{pa}X^{rd}\lambda_{a}%
(\lambda^{2})_{,p}\ .\label{Q1Q2}%
\end{align}
Besides,%
\begin{equation}
\mathrm{Str}\left(  V_{2}^{2}\right)  =\mathrm{Str}^{2}\left(  V_{2}\right)
\not \equiv 0\ .\label{strq22}%
\end{equation}
Indeed,%
\begin{align}
&  \ \left(  V_{2}\right)  _{p}^{p}\left(  -1\right)  ^{\varepsilon_{p}%
}=\lambda_{a}Y^{p}\lambda_{,p}^{a}\ ,\label{indeed1}\\
&  \ \left(  Q_{2}\right)  _{q}^{p}\left(  Q_{2}\right)  _{p}^{q}\left(
-1\right)  ^{\varepsilon_{p}}=\left(  \lambda_{a}Y^{p}\lambda_{,p}^{a}\right)
\left(  \lambda_{b}Y^{q}\lambda_{,q}^{b}\right)  \ .\label{indeed2}%
\end{align}
Therefore, $\Im$ in the expression (\ref{superJ}) for the Jacobian
$\exp\left(  \Im\right)  $ has the general structure%
\begin{align}
&  \Im=A\left(  f,V_{1},W\right)  +B\left(  f|V_{2}\right)  +C\left(
f|V_{1}V_{2}\right)  \ ,\label{strIm}\\
&  \mathrm{for}\,\,\,B\left(  f|V_{2}\right)  =b_{1}\left(  f\right)
\mathrm{Str}\left(  V_{2}\right)  +b_{2}\left(  f\right)  \mathrm{Str}\left(
V_{2}^{2}\right)  =\left[  b_{1}\left(  f\right)  +b_{2}\left(  f\right)
\mathrm{Str}\left(  V_{2}\right)  \right]  \mathrm{Str}\left(  V_{2}\right)
\ ,\nonumber\\
&  \mathrm{and}\,\,\,C\left(  f|V_{1}V_{2}\right)  =c\left(  f\right)
\mathrm{Str}\left(  V_{1}V_{2}\right)  \ .\nonumber
\end{align}
Let us examine $A\left(  f,V_{1},W\right)  $. Namely, in view of (\ref{zero})
and $\mathrm{Str}\left(  U^{n}\right)  =-2f^{n}$, we have%
\begin{align}
A\left(  f,V_{1},W\right)   &  =\mathrm{Str}\left(  V_{1}\right)
+\mathrm{Str}\left(  W\right)  -\frac{1}{2}\mathrm{Str}\left(  V_{1}%
^{2}\right)  +2\sum_{n=1}^{\infty}\frac{\left(  -1\right)  ^{n}}{n}%
f^{n}\label{Af}\\
&  =-\left(  \Delta^{a}S\right)  \lambda_{a}-\frac{1}{4}\left(
\Delta ^{a}S\right)  \overleftarrow{s}_{a}\lambda^{2}-2\ln\left(
1+f\right) \,.\nonumber
\end{align}
Let us examine the explicit structure of the series related to $b_{1}\left(
f\right)  $: the quantity $\mathrm{Str}\left(  V_{2}\right)  $ derives from
$\mathrm{Str}\left(  U^{n-1}V\right)  $ for $n\geq1$ in (\ref{strPnQk}), and
is coupled with the combinatorial coefficient $C_{n}^{1}$. The part of $\Im$
containing $\mathrm{Str}\left(  V_{2}\right)  $ is given by%
\begin{equation}
b_{1}\left(  f\right)  \mathrm{Str}\left(  V_{2}\right)  =C_{1}^{1}%
\mathrm{Str}\left(  V_{2}\right)  -\sum_{n=2}^{\infty}\frac{\left(  -1\right)
^{n}}{n}C_{n}^{1}f^{n-2}\left(  1+f\right)  \mathrm{Str}\left(  V_{2}\right)
\ ,\label{b1f}%
\end{equation}
whence%
\begin{equation}
b_{1}\left(  f\right)  =1-\left(  1+f\right)  \sum_{n=0}^{\infty}\left(
-1\right)  ^{n}f^{n}
\,.\label{b1ffin}%
\end{equation}
Let us examine the explicit structure of the series related to $b_{2}\left(
f\right)  $: the quantity $\mathrm{Str}^{2}\left(  V_{2}\right)  $ derives
from $\mathrm{Str}\left(  U^{n-2}V^{2}\right)  $ for $n\geq2$ in
(\ref{strPnQk}), coupled with the combinatorial coefficients $C_{n}^{2}$ for
$n=2,3$ and $\left(  C_{n}^{2}-K_{n}\right)  $ for $n>3$, and also derives
from $\mathrm{Str}\left(  U^{n-3}VUV\right)  $ for $n>3$ in (\ref{strPnQk}),
coupled with the combinatorial coefficients $K_{n}$. The part of $\Im$
containing $\mathrm{Str}^{2}\left(  V_{2}\right)  $ reads%
\begin{align}
b_{2}\left(  f\right)  \mathrm{Str}^{2}\left(  V_{2}\right)  = &
-\frac{\left(  -1\right)  ^{2}}{2}C_{2}^{2}\mathrm{Str}^{2}\left(
V_{2}\right)  -\frac{\left(  -1\right)  ^{3}}{3}C_{3}^{2}\left(  1+f\right)
\mathrm{Str}^{2}\left(  V_{2}\right)  \nonumber\\
&  -\sum_{n=4}^{\infty}\frac{\left(  -1\right)  ^{n}}{n}\left(  C_{n}%
^{2}-K_{n}\right)  f^{n-3}\left(  1+f\right)  \mathrm{Str}^{2}\left(
V_{2}\right)  -\sum_{n=4}^{\infty}\frac{\left(  -1\right)  ^{n}}{n}%
K_{n}f^{n-4}\left(  1+f\right)  ^{2}\mathrm{Str}^{2}\left(  V_{2}\right)
\ ,\label{b2f}%
\end{align}
whence%
\begin{align}
b_{2}\left(  f\right)   &  =-\frac{1}{2}+\left(  1+f\right)  -\sum
_{n=4}^{\infty}\frac{\left(  -1\right)  ^{n}}{n}\left[  \left(  C_{n}%
^{2}-K_{n}\right)  f^{n-3}\left(  1+f\right)  +K_{n}f^{n-4}\left(  1+f\right)
^{2}\right]
\ .\label{b2ffin}%
\end{align}
Let us examine the explicit structure of the series related to $c\left(
f\right)  $: the quantity $\mathrm{Str}\left(  V_{1}V_{2}\right)  $ derives
from $\mathrm{Str}\left(  U^{n-2}V^{2}\right)  $ for $n\geq2$ in
(\ref{strPnQk}), and is coupled with the combinatorial coefficients $C_{n}%
^{2}$, for $n=2,3$, and $C_{n}^{2}-K_{n}$, for $n>3$. The part of $\Im$
containing $\mathrm{Str}\left(  V_{1}V_{2}\right)  $ is given by%
\begin{align}
c\left(  f\right)  \mathrm{Str}\left(  V_{1}V_{2}\right)  = &  -\frac{\left(
-1\right)  ^{2}}{2}C_{2}^{2}\mathrm{Str}\left(  2V_{1}V_{2}\right)
-\frac{\left(  -1\right)  ^{3}}{3}C_{3}^{2}\left(  1+f\right)  \mathrm{Str}%
\left(  V_{1}V_{2}\right)  \nonumber\\
&  -\sum_{n=4}^{\infty}\frac{\left(  -1\right)  ^{n}}{n}\left(  C_{n}%
^{2}-K_{n}\right)  f^{n-3}\left(  1+f\right)  \mathrm{Str}\left(  V_{1}%
V_{2}\right)  \ ,\label{cf_ini}%
\end{align}
whence%
\begin{equation}
c\left(  f\right)  =-1+\left(  1+f\right)  -\sum_{n=4}^{\infty}\frac{\left(
-1\right)  ^{n}}{n}\left(  C_{n}^{2}-K_{n}\right)  f^{n-3}\left(  1+f\right)
=f-\left(  1+f\right)  \sum_{n=4}^{\infty}\left(  -1\right)  ^{n}\left(
\frac{C_{n}^{2}}{n}-\frac{K_{n}}{n}\right)  f^{n-3}\ .\label{cf}%
\end{equation}
By virtue of (\ref{C,K}), one can show \cite{MRnew} that
the series $b_{1}\left(  f\right) $, $b_{2}\left(  f\right)  $, $c\left(  f\right)  $,
given by (\ref{b1ffin}), (\ref{b2ffin}), (\ref{cf_ini}),
vanish identically:
\begin{equation}
b_{1}\left(  f\right)=b_{2}\left(  f\right)=c\left(  f\right)\equiv 0\,.
\label{bc}
\end{equation}
From the vanishing of all the coefficients $b_{1}\left(  f\right)  $,
$b_{2}\left(  f\right)  $, $c\left(  f\right)  $,
we conclude that%
\begin{equation}
B\left(  f|V_{2}\right)  =b_{1}\left(  f\right)  \mathrm{Str}\left(
V_{2}\right)  +b_{2}\left(  f\right)  \mathrm{Str}\left(  V_{2}^{2}\right)
\equiv0\,\,\,\,\mathrm{and}\,\,\,C\left(  f|V_{1}V_{2}\right)  =c\left(
f\right)  \mathrm{Str}\left(  V_{1}V_{2}\right)  \equiv0\ ,\label{BCf}%
\end{equation}
and therefore the Jacobian $\exp\left(  \Im\right)  $ is finally given by%
\begin{align*}
&  \ \Im=A\left(  f,V_{1},W\right)  +B\left(  f|V_{2}\right)  +C\left(
f|V_{1}V_{2}\right)  =A\left(  f,V_{1},W\right)  =-\left(  \Delta^{a}S\right)
\lambda_{a}-\frac{1}{4}\left(  \Delta^{a}S\right)  \overleftarrow{s}%
_{a}\lambda^{2}-2\ln\left(  1+f\right)  \ ,\\
&  \ \mathrm{for}\,\,\,f=-\left(  1/2\right)  \Lambda\overleftarrow{U}{}%
^{2}\,=-\left(  1/2\right)  \Lambda\overleftarrow{s}{}^{2}\ ,\ \ \ \Lambda
=\Lambda\left(  \phi,\pi,\lambda\right)  \ .
\end{align*}
which is identical with (\ref{superJaux}) and therefore proves (\ref{superJ1}).

\section{Ward Identities and Gauge Dependence Problem}

\label{WIGD}\renewcommand{\theequation}{\arabic{section}.\arabic{equation}}
\setcounter{equation}{0} In this section, we touch upon the
consequences (Subsection~\ref{WIGD1}) implied by a solution of the
compensation equation \cite{MRnew2} for an unknown functional
$\Lambda(\phi,\pi,\lambda)$ which determines a field-dependent
BRST-antiBRST transformation that amounts to a precise change of the
gauge-fixing functional for an arbitrary gauge theory.
The modified Ward identities and gauge dependence
for Yang--Mills theories and, more generally, for gauge
theories with a closed algebra of rank 1 are examined in
Subsection~\ref{WIGD2}.

\subsection{General Gauge Theory}

\label{WIGD1} Let us bring to mind the results of \cite{MRnew2}, based on
the  representation (\ref{superJ1}) for the Jacobian
established in Section~\ref{gensetup} for the change of variables
(\ref{Gamma_fin}), given by a field-dependent BRST-antiBRST
transformation, $\Gamma\rightarrow\check{\Gamma}=\Gamma+\Delta_\lambda\Gamma$.
First of all, making in $Z_{F}$ a change of variables
$\Gamma\rightarrow\check{\Gamma}$, with $\lambda_{a} = s_{a}\Lambda$ and
$\Lambda=\Lambda(\phi, \pi,\lambda)$,
while taking account of the Jacobian (\ref{superJ1}), the change of the
action $\mathcal{S}_{F}$ according to (\ref{DeltaF}),
and also using the invariance condition (\ref{eqinv}), related to the generating equations (\ref{3.3}),
and its consequence resulting from applying the operator
$\overleftarrow{s}_{a}$, with allowance made for the BRST-antiBRST invariance
of the term $F\overleftarrow{U}{}^{2}$, we arrive at
\begin{eqnarray}
 Z_{F} & \stackrel{\Gamma\rightarrow
\check{\Gamma}}{=}& \int d\Gamma \;\exp\left\{  \frac{i}{\hbar}  \left[
\mathcal{S}_{F}  +  \left( {\cal S}_{F} \overleftarrow{s}^a + {i}{\hbar
}\Delta^{a}S\right)  \lambda_{a}+\frac{1}{4}\left({\cal S}_{F} \overleftarrow{s}{}^2 + {i}{\hbar
} \Delta
^{a}S  \overleftarrow{s}_{a}\right)\lambda^{2}+i\hbar\,\mathrm{\ln}\left(
1-\frac{1}{2}\Lambda\overleftarrow{s}{}^{2}\right)  ^{2}  \right]  \right\} \nonumber \\
   &=&  \int d\Gamma \;\exp\left\{  \frac{i}{\hbar}  \left[
\mathcal{S}_{F+\Delta F}  + i\hbar\,\mathrm{\ln}\left(
1-\frac{1}{2}\Lambda\overleftarrow{s}{}^{2}\right)  ^{2}
+\frac{1}{2}\Delta F\overleftarrow{U}{}^{2}  \right]  \right\}\,.
\label{zftrans}
\end{eqnarray}
The coincidence of the vacuum functionals $Z_{F}$ and $Z_{F+\Delta
F}$, evaluated for the respective Bosonic functionals $F$ and
$F+\Delta F$, takes place in case there holds a compensation
equation\footnote{In \cite{MRnew2}, we make a transformation
parameterized by $\Lambda$ to perform a transition from
$Z_{F+\Delta F}$ to $Z_{F}$, which accounts for the opposite sign
at $\Delta F$ in (\ref{eqexpl}) and related formulae.} for an
unknown Bosonic functional $\Lambda=\Lambda(\phi,\pi,\lambda)$:
\begin{equation}
i\hbar\,\mathrm{\ln}\left(
1-\frac{1}{2}\Lambda\overleftarrow{U}{}^{2}\right)  ^{2}
=-\frac{1}{2}\Delta F\overleftarrow{U}{}^{2}\,,
\label{eqexpl}%
\end{equation}
where account has been taken of the relation
$\overleftarrow{s}{}^{a} = \overleftarrow{U}{}^{a}$, which takes
place for the operators $\overleftarrow{s}{}^{a}$ restricted to
the variables $(\phi,\pi^a,\lambda)$, or, equivalently,
\begin{equation}
\frac{1}{2}\Lambda\overleftarrow{U}{}^{2}=1-\exp\left(  {{\frac{i}{\,4\hbar}%
}\Delta F\overleftarrow{U}{}^{2}}\right)  \ . \label{eqexpllam}%
\end{equation}
The solution of this equation for an unknown functional $\Lambda\left(
\phi,\pi,\lambda\right)  $, which determines $\lambda_{a}\left(  \phi
,\pi,\lambda\right)  $ according to $\lambda_{a}=\Lambda\overleftarrow
{U}_{a}$, with accuracy up to BRST-antiBRST exact terms ($s^{a}$ being restricted to
$\phi$, $\pi^{a}$, $\lambda$), is given by%
\begin{equation}
\Lambda(\Gamma|\Delta{F})=\frac{i}{2\hbar}g(y)\Delta{F}\ ,\ \ \mathrm{for}%
\ \ g(y)=\left[  1-\exp(y)\right]  /y\ \ \mathrm{and}\ \ y\equiv\frac
{i}{4\hbar}\Delta F\overleftarrow{U}{}^{2}\ , \label{solcompeq2}%
\end{equation}
and therefore the corresponding field-dependent parameters have the form%
\begin{equation}
\lambda_{a}\left(  \Gamma|\Delta{F}\right)  =\frac{i}{2\hbar}g(y)\left(
\Delta{F}\overleftarrow{U}_{a}\right)  \, , \label{funcdeplafin}%
\end{equation}
whose approximation linear in $\Delta{F}$ is given by
\begin{equation}
\lambda_{a}\left(  \Gamma|\Delta{F}\right)  = \frac{i}{2\hbar}\left(
\Delta{F}\overleftarrow{U}_{a}\right)  + {o}(\Delta{F}) \,.
\label{funcdeplainf}%
\end{equation}
Consequently, for any change $\Delta F$ of the gauge condition $F\to F+ \Delta F$,
we can construct a unique field-dependent BRST-antiBRST transformation
with functionally-dependent parameters (\ref{funcdeplafin}) that allows one to
preserve the form of the partition function (\ref{zftrans}) for the same gauge
theory. On the other hand, if we consider the compensation equation (\ref{eqexpl})
for an unknown gauge variation $\Delta F$ with a given $\Lambda\left(  \phi,\pi
,\lambda\right) $, we can present it in the form
\begin{equation}
4i\hbar\ln\left(  1-\frac{1}{2}\Lambda\overleftarrow{U}{}^{2}\right)
=- \Delta F\overleftarrow{U}{}^{2}\ \Longleftrightarrow\ 4i\hbar\left[
\sum_{n=1}\frac{(-1)^{n}}{2^{n} n}\left( \Lambda\overleftarrow{U}{}^{2}\right)
{}^{n-1} \Lambda\right] \overleftarrow{U}{}^{2} = \Delta F\overleftarrow{U}%
{}^{2}\,, \label{eqexp2}%
\end{equation}
whose solution, with accuracy up to $\overleftarrow{U}{}^{a}$-exact terms,
is given by
\begin{equation}
\label{oppsol}\Delta F(\Gamma| \Lambda) = 4i\hbar\left[ \sum
_{n=1}\frac{(-1)^{n}}{2^{n} n}\left( \Lambda\overleftarrow{U}{}^{2}\right)
{}^{n-1} \Lambda\right]  = 4i\hbar\left[ \sum_{n=1}\frac{(-1)^{n}%
}{2^{n} n}\left( \lambda^{a}\overleftarrow{U}_{a}\right) {}^{n-1}
\Lambda\right] .
\end{equation}
Consequently, for any change of variables in the partition function $Z_{F}$
given by finite field-dependent BRST-antiBRST transformations with the parameters
$\lambda^{a} = \Lambda\overleftarrow{U}{}^{a}$, we obtain
the same partition function $Z_{F+\Delta F}$,
however, evaluated in a gauge determined by the Bosonic functional $F+\Delta F$,
in accordance with (\ref{oppsol}).

Making in (\ref{z(0)}) a field-dependent BRST-antiBRST transformation
(\ref{Gamma_fin}) and using the relations (\ref{superJ1}) and (\ref{solcompeq2}),
we obtain a \emph{modified Ward} (\emph{Slavnov--Taylor}%
) \emph{identity:}%
\begin{equation}
\left\langle \left\{  1+\frac{i}{\hbar}J_{A}\phi^{A}\left[  \overleftarrow
{U}^{a}\lambda_{a}(\Lambda)+\frac{1}{4}\overleftarrow{U}^{2}\lambda
^{2}(\Lambda)\right]  -\frac{1}{4}\left(  \frac{i}{\hbar}\right)  {}^{2}%
J_{A}\phi^{A}\overleftarrow{U}^{a}J_{B}(\phi^{B})\overleftarrow{U}_{a}%
\lambda^{2}(\Lambda)\right\}  \left(  1-\frac{1}{2}\Lambda\overleftarrow
{U}^{2}\right)  {}^{-2}\right\rangle _{F,J} =1\ . \label{mWI}%
\end{equation}
Here, the symbol \textquotedblleft$\langle\mathcal{O}\rangle_{F,J}%
$\textquotedblright\ for a quantity $\mathcal{O}=\mathcal{O}(\Gamma)$ stands
for a source-dependent average expectation value corresponding to a
gauge-fixing $F(\phi)$:
\begin{equation}
\left\langle \mathcal{O}\right\rangle _{F,J}=Z_{F}^{-1}(J)\int d\Gamma
\ \mathcal{O}\left(  \Gamma\right)  \exp\left\{  \frac{i}{\hbar}\left[
\mathcal{S}_{F}\left(  \Gamma\right)  +J_{A}\phi^{A}\right]  \right\}
\ ,\ \ \mathrm{with\ \ }\left\langle 1\right\rangle _{F,J}=1\ . \label{aexv}%
\end{equation}
Owing to the presence of $\Lambda(\Gamma)$, which implies $\lambda_{a}(\Lambda
)$, the modified Ward identity depends on a choice of the gauge Boson
$F(\phi)$ for non-vanishing $J_{A}$, according to (\ref{solcompeq2}),
(\ref{funcdeplafin}), and therefore the corresponding Ward identities for
Green's functions, obtained by differentiating (\ref{mWI}) with respect to the
sources, contain the functionals $\lambda_{a}(\Lambda)$ and their derivatives
as weight functionals. By virtue of (\ref{mWI}) for constant $\lambda_{a}$,
the usual Ward identities follow from the first order in $\lambda_{a}$,
and a new Ward identity \cite{MRnew2} from the second order in $\lambda_{a}$:
\begin{equation}
J_{A}\left\langle \phi^{A}\overleftarrow{U}{}^{a}\right\rangle _{F,J}=0,\quad
\left\langle J_{A}\phi^{A}\left[  \overleftarrow{U}{}^{2}-\overleftarrow
{s}^{a}\left(  i/\hbar\right)  J_{B}\left(  \phi^{B}\overleftarrow{U}%
_{a}\right)  \right]  \right\rangle _{F,J}=0\ . \label{WIlag2}%
\end{equation}
In conclusion, taking account of (\ref{funcdeplafin}),
we find that (\ref{mWI}) implies a relation which describes
the gauge dependence of $Z_{F}(J)$ for a finite change
of the gauge, $F\rightarrow F+\Delta F$:%
\begin{align}
Z_{F+\Delta F}(J)  &  =Z_{F}(J)\left\{  1+\left\langle \frac{i}{\hbar}%
J_{A}\phi^{A}\left[  \overleftarrow{U}^{a}\lambda_{a}\left(  \Gamma|-\Delta
{F}\right)  +\frac{1}{4}\overleftarrow{U}^{2}\lambda^{2}\left(  \Gamma
|-\Delta{F}\right)  \right]  \right. \right. \nonumber\\
&  -\left.  \left.  (-1)^{\varepsilon_{B}}\left(  \frac{i}{2\hbar}\right)
^{2}J_{B}J_{A}\left(  \phi^{A}\overleftarrow{U}{}^{a}\right)  \left(  \phi
^{B}\overleftarrow{U}_{a}\right)  \lambda^{2}\left(  \Gamma|-\Delta{F}\right)
\right\rangle _{F,J} \right\}  . \label{GDInew}%
\end{align}

\subsection{Yang--Mills Theory}

\label{WIGD2}

Let us apply the modified Ward identities (\ref{mWI}) and the
representation (\ref{GDInew}) for gauge dependence to the
generating functional $Z_{F}(J)$ of irreducible gauge theories of
rank 1 with a closed gauge algebra of the generators of gauge
transformations, $R_{\alpha}^{i}(A)$,
$\varepsilon(R_{\alpha}^{i})= \varepsilon_i + \varepsilon_\alpha$,
including the case of Yang--Mills theories. The corresponding
configuration space $\phi^A = \left( A^{i},B^{\alpha},C^{\alpha
a}\right)$ contains the classical fields, the Nakanishi--Lautrup
fields, and the ghost-antighost fields \cite{BLT1}. First of all,
in accordance with \cite{MRnew}, the generating functional of
Green's functions $Z_{F}(J)$ and the corresponding quantum action
$S_F(\phi)$ are readily obtained from (\ref{z(0)}), with $S$ being
a solution of (\ref{3.3}), by integrating over $(\phi^*_a,
\bar{\phi},\pi^a,\lambda)$:
\begin{align}
Z_{F}(J)  &  =  \int d\phi\ \exp\left\{  \frac{i}{\hbar}\left[  S_{F}\left(
\phi\right)  +J_{A}\phi^{A}\right]  \right\} ,\label{zj}\\
S_{F}\left(  \phi\right)   &  =  S_{0}\left(  A\right) - \left(1/2\right) F\overleftarrow
{s}{}^{2} = S_{0}\left(  A\right)  +F_{,A}Y^{A}-\left(  1/2\right)
\varepsilon_{ab}X^{Aa}F_{,AB}X^{Bb}\ , \label{action}%
\end{align}
where
\begin{align}
&  X^{Aa}=\left(  X_{1}^{ia},X_{2}^{\alpha a},X_{3}^{\alpha ab}\right)  \ , &
&  Y^{A}=\left(  Y_{1}^{i},Y_{2}^{\alpha},Y_{3}^{\alpha a}\right)
\ ,\label{solxy}\\
&  X_{1}^{ia}=R_{\alpha}^{i}C^{\alpha a}\ , &  &  X_{2}^{\alpha a}=-\frac
{1}{2}F_{\gamma\beta}^{\alpha}B^{\beta}C^{\gamma a}-\frac{1}{12}\left(
-1\right)  ^{\varepsilon_{\beta}}\left(  2F_{\gamma\beta,j}^{\alpha}R_{\rho
}^{j}+F_{\gamma\sigma}^{\alpha}F_{\beta\rho}^{\sigma}\right)  C^{\rho
b}C^{\beta a}C^{\gamma c}\varepsilon_{cb}\ ,\nonumber\\
&  X_{3}^{\alpha ab}=-\varepsilon^{ab}B^{\alpha}-\frac{1}{2}\left(  -1\right)
^{\varepsilon_{\beta}}F_{\beta\gamma}^{\alpha}C^{\gamma b}C^{\beta a}\ , &  &
Y_{1}^{i}=R_{\alpha}^{i}B^{\alpha}+\frac{1}{2}\left(  -1\right)
^{\varepsilon_{\alpha}}R_{\alpha,j}^{i}R_{\beta}^{j}C^{\beta b}C^{\alpha
a}\varepsilon_{ab}\ ,\nonumber\\
&  Y_{2}^{\alpha}=0\ , &  &  Y_{3}^{\alpha a}=-2X_{3}^{\alpha a}\ , \label{xy}%
\end{align}
and there hold the subsidiary conditions $X^{Aa}_{,A}=0$.
The nilpotent, $\overleftarrow{s}{}^{a}\overleftarrow{s}{}^{b}\overleftarrow{s}%
{}^{c}\equiv0$, generators of BRST-antiBRST transformations $\overleftarrow
{s}{}^{a}$ determine the finite BRST-antiBRST transformations:
\begin{equation}
\Delta_\lambda\phi^{A}=X^{Aa}\lambda_{a}-\frac{1}{2}Y^{A}\lambda^{2}=\phi^{A}\left(
\overleftarrow{s}{}^{a} \lambda_{a}+\frac{1}{4} \overleftarrow{s}{}^{2}%
\lambda^{2}\right)  \ , \label{finite}%
\end{equation}
so that the corresponding Jacobian $\mathrm{Sdet}\left\Vert \left(  \phi
^{A}+\Delta_\lambda\phi^{A}\right)  \frac{\overleftarrow{\delta}}{\delta\phi^{B}%
}\right\Vert $ of a change of variables $\phi\to \phi\left(  1 +
\overleftarrow{s}{}^{a} \lambda_{a}+\frac{1}{4} \overleftarrow
{s}{}^{2}\lambda^{2}\right)  $ takes the form ($X^{Aa}_{,A}=0$)
\begin{align}
&\mathrm{Sdet}\left\Vert \left(
\phi^{A}+\Delta_\lambda\phi^{A}\right)
\frac{\overleftarrow{\delta}}{\delta\phi^{B}}\right\Vert =
\exp\left[
X_{,A}^{Aa}\lambda_{a}+\frac{1}%
{4}\varepsilon_{ab}X_{,AB}^{Aa}X^{Bb}\lambda^{2}
+\ln\left(  1-\frac{1}{2}\Lambda\overleftarrow{s}{}^{2}\right)
^{-2}\right]
\nonumber\\
&=\exp\left[\ln\left(  1-\frac{1}{2}\Lambda\overleftarrow{s}{}^{2}\right)
^{-2}\right]
\ ,\ \mathrm{ for }\ \lambda_{a} = \Lambda\overleftarrow{s}{}_{a}\,,%
\label{superJaux1}\\
&\mathrm{Sdet}\left\Vert \left(  \phi^{A}+\Delta_\lambda\phi^{A}\right)
\frac{\overleftarrow{\delta}}{\delta\phi^{B}}\right\Vert
=\exp\left(
X_{,A}^{Aa}\lambda_{a}+\frac{1}%
{4}\varepsilon_{ab}X_{,AB}^{Aa}X^{Bb}\lambda^{2}\right)  = 1  \ ,\ \mathrm{ for
}\ \lambda_{a} =\mathrm{ const } \ . \label{superJ2}%
\end{align}
The conditions $X^{Aa}_{,A} =0$ imply the relations $R_{\alpha, i}^{i}(A) = F_{\alpha\beta}^{\alpha}(A)=0$
for the gauge generators and structure functions in $R_{\alpha,j}^{i}(A)R_{\beta}^{j}(A)-\left(  -1\right)
^{\varepsilon_{\alpha}\varepsilon_{\beta}}R_{\beta,j}^{i}(A)R_{\alpha}%
^{j}(A)=-R_{\gamma}^{i}(A)F_{\alpha\beta}^{\gamma}\left(  A\right)  $.
The first Jacobian (\ref{superJaux1}), corresponding to
field-dependent (and functionally-dependent) parameters
$\lambda_{a}(\phi) =  (\Lambda\overleftarrow{s}{}_a)(\phi) $,
is identical with that of \cite{MRnew}.

As we apply the procedure used to obtain (\ref{mWI}) to
gauge theories with a closed algebra of rank 1, we arrive at a \emph{modified Ward identity}:
\begin{equation}
\left\langle \left\{  1+\frac{i}{\hbar}J_{A}\left[ X^{Aa} \lambda_{a}%
(\Lambda)-\frac{1}{2}Y^{A}\lambda^{2}(\Lambda)\right]  -\frac{1}{4}\left(
\frac{i}{\hbar}\right)  {}^{2}\varepsilon_{ab}J_{A}X^{Aa}J_{B}X^{Bb}%
\lambda^{2}(\Lambda)\right\}  \left(  1-\frac{1}{2}\Lambda\overleftarrow
{s}^{2}\right)  {}^{-2}\right\rangle _{F,J} =1\ , \label{mWIclalg}%
\end{equation}
where the symbol \textquotedblleft$\langle\mathcal{O}\rangle_{F,J}%
$\textquotedblright\ for a quantity $\mathcal{O}=\mathcal{O}(\phi)$ is
determined as in (\ref{aexv}), however, for $Z_{F}(J)$ in (\ref{zj}).
The identity (\ref{mWIclalg}) has the same interpretation as
(\ref{mWI}): for constant $\lambda_{a}$ we obtain from (\ref{mWIclalg})
an Sp(2)-doublet of the usual Ward identities at the first order in
$\lambda_{a}$, and a derivative identity at the second
order in $\lambda_{a}$:
\begin{equation}
J_{A}\left\langle X^{Aa}\right\rangle _{F,J}=0,\quad  \left\langle J_{A}\left[ 2Y^{A}
+\left(  i/\hbar\right)\varepsilon_{ab} X^{Aa}  J_{B} X^{Bb}
\right]  \right\rangle _{F,J}=0\ . \label{WIlag3}%
\end{equation}
By virtue of the representation (\ref{funcdeplafin}) for $\lambda_{a}(\phi| \Delta F)$, applied to gauge theories
in question, the Ward identity (\ref{mWIclalg}) implies a relation which describes the gauge
dependence of $Z_{F}(J)$ for a finite change of the gauge $F\rightarrow
F+\Delta F$:%
\begin{align}
Z_{F+\Delta F}(J) - Z_{F}(J)  &  =Z_{F}(J)\left\langle \frac{i}{\hbar}%
J_{A}\left[ X^{Aa} \lambda_{a}\left(  \phi|-\Delta{F}\right)  -\frac{1}%
{2}Y^{A}\lambda^{2}\left(  \phi|-\Delta{F}\right)  \right]  \right.
\nonumber\\
&  - \left.  (-1)^{\varepsilon_{B}}\left(  \frac{i}{2\hbar}\right)  ^{2}%
J_{B}J_{A}\left(  X^{Aa}X^{Bb}\right)  \varepsilon_{ab} \lambda^{2}\left(
\phi|-\Delta{F}\right) \right\rangle _{F,J} . \label{GDInew1}%
\end{align}
For Yang--Mills theories, in which $X^{Aa}_{,A}\equiv 0$,
we obtain a new representation for the
\emph{modified Ward identity} (\ref{mWIclalg}), with the following
identification of $X^{Aa}$ and $Y^{A}$ in (\ref{xy}), according to \cite{MRnew,RML}:
\begin{align}
&  X_{1}^{\mu ma}=D^{\mu mn}C^{na}\ , &  &  Y_{1}^{\mu m}=D^{\mu mn}%
B^{n}+\frac{1}{2}f^{mnl}C^{la}D^{\mu nk}C^{kb}\varepsilon_{ba}\ ,\nonumber\\
&  X_{2}^{ma}=-\frac{1}{2}f^{mnl}B^{l}C^{na}-\frac{1}{12}f^{mnl}f^{lrs}%
C^{sb}C^{ra}C^{nc}\varepsilon_{cb}\ , &  &  Y_{2}^{m}=0\ ,\label{xyYM}\\
&  X_{3}^{mab}=-\varepsilon^{ab}B^{m}-\frac{1}{2}f^{mnl}C^{lb}C^{na}\ , &  &
Y_{3}^{ma}=f^{mnl}B^{l}C^{na}+\frac{1}{6}f^{mnl}f^{lrs}C^{sb}C^{ra}%
C^{nc}\varepsilon_{cb}\ ,\nonumber
\end{align}
where $\phi^{A}=\left(  A^{m\mu},B^{m},C^{ma}\right) $; the
generators $R_{\mu}^{mn}(x;y)$ of gauge transformations, the
covariant derivatives $D_{\mu}^{mn}$ and the structure functions
$F_{\alpha\beta}^{\gamma}$ are given by \cite{MRnew,RML}
\begin{equation}
R_{\mu}^{mn}(x;y) = D_{\mu}^{mn}(x)\delta(x-y)\, ,\quad D_{\mu}^{mn}=\delta
^{mn}\partial_{\mu}+f^{mln}A_{\mu}^{l}\,,\quad F_{\alpha\beta}^{\gamma}%
=f^{lmn}\delta(x-z)\delta(y-z)\,. \label{R(A)}%
\end{equation}

\section{BRST-antiBRST Symmetry Breaking in Gauge Theories}

\label{BaBsb}\renewcommand{\theequation}{\arabic{section}.\arabic{equation}} \setcounter{equation}{0}

In this section, we introduce the concept of \emph{BRST-antiBRST
symmetry breaking} in gauge theories, inspired by our research
\cite{llr1} within the BV quantization scheme and  partially
repeating the study of \cite{Reshetnyak} in which the problem of
inconsistency of BRST symmetry breaking introduction claimed in
\cite{llr1} was  resolved. To this end, let us consider a Bosonic
functional $M = M(\phi, \phi^*, \bar{\phi})$ having a vanishing
ghost number and not necessarily being BRST-antiBRST invariant.
This functional may be added to the action $\mathcal{S}_F$ in
(\ref{action}), or introduced in a multiplicative way into the
partition function, $m = \exp{(M)}$, in order to improve the
properties of the path integral, as has been done within the
functional renormalization group approach
\cite{Reshetnyak,Wett-Reu-1, Wett-Reu-2, LS, Polch}, by means of
the average affective action for the purpose of extraction of
residual Gribov copies \cite{Gribov,Zwanziger}; see also
\cite{Reshetnyak2, sorellas, sorellas2}. Let us impose the
condition that the corresponding path integral be well-defined and
determine the generating functional of Green's functions
$Z_{M,F}(J)$ with \emph{broken BRST-antiBRST symmetry} as follows:
\begin{align}
&  Z_{M,F}(J)=\int d\Gamma\;\exp\left\{  \left(  i/\hbar\right)  \left[
\mathcal{S}_{F}\left(  \Gamma\right)+  M(\phi, \phi^*, \bar{\phi}) +J_{A}\phi^{A}\right]  \right\}
\ ,\ \ \mathrm{ with } \ \ Z_{0,F}(J)\equiv Z_{F}(J) .\label{z(0)m}%
\end{align}
Here, we have not imposed\footnote{On a basis of this equation,
one can develop a concept of soft BRST-antiBRST symmetry breaking
in gauge theories, which may lead to additional Ward identities in
terms of the functional $M$; however, we leave this problem
outside the scope of the present work.} on $M$ any equation
($\frac{1}{2}(M,M)^a - V^a M = 0 $ or $\frac{1}{2}(M,M)^a - V^a M
= -i\hbar \Delta^a M $) analogous to the so-called \emph{soft BRST
symmetry equation} \cite{llr1,lrr} for gauge theories in the BV
formalism. Another requirement for  $M$ is the following
inequality, in terms of the operators $\overleftarrow{s}{}^a$ in
(\ref{Gamma_fin}), with account taken of the representation
(\ref{DeltaF}) for a finite BRST-antiBRST variation of an
arbitrary functional:
\begin{equation}\label{notBab}
  M\overleftarrow{s}{}^a \ne 0 \quad  \Longrightarrow \quad \Delta_\lambda M= M \left( \overleftarrow{s}{}^{a}\lambda_{a}+\frac{1}{4}%
\overleftarrow{s}{}^{2}\lambda^{2}\right)  \ne 0\,.
\end{equation}
We shall refer to a gauge theory as having a broken BRST-antiBRST symmetry
in the Sp(2)-covariant Lagrangian quantization \cite{BLT1,BLT2}
if the total action of the theory, $\mathcal{S}_{\mathrm{tot}} = \mathcal{S}_{F} +  M$,
determines the partition function $Z_{M,F}(0)$ in (\ref{z(0)m})
and the broken BRST-antiBRST symmetry condition (\ref{notBab}) for $M$ is fulfilled.

It is well known that the introduction of a BRST-antiBRST non-invariant term
to the quantum action may lead to the appearance of two fundamental problems
for physical quantities (such as the S-matrix) in a theory with
broken BRST-antiBRST symmetry, namely, gauge dependence and unitarity failure.
For this reason, let us study the suggested construction
of a gauge theory with broken BRST-antiBRST symmetry
as concerns the dependence of the functional $Z_{M,F}(J)$
on a choice of the gauge condition.

Let us first obtain the  Ward identities for a gauge theory with broken BRST-antiBRST symmetry.
To this end, we make in (\ref{z(0)m}) a field-dependent BRST-antiBRST transformation
(\ref{Gamma_fin}) and use the relations (\ref{solcompeq2}) and (\ref{superJ1})
to obtain a \emph{modified Ward} (\emph{Slavnov--Taylor} for $ Z_{M,F}(J)$%
) \emph{identity:}%
\begin{eqnarray}
&& \left\langle \left\{  1+\frac{i}{\hbar}\left[J_{A}\phi^{A}+M_F\right]\left[  \overleftarrow
{U}^{a}\lambda_{a}(\Lambda)+\frac{1}{4}\overleftarrow{U}^{2}\lambda
^{2}(\Lambda)\right]  -\frac{1}{4}\left(  \frac{i}{\hbar}\right)  {}^{2}%
\left[J_{A}\phi^{A}+M_F\right]\overleftarrow{U}^{a}\left[J_{B}\phi^{B}+M_F\right]\overleftarrow{U}_{a}%
\lambda^{2}(\Lambda)\right\}\right.\nonumber\\
&& \qquad \left. \times\left(  1-\frac{1}{2}\Lambda\overleftarrow
{U}^{2}\right)  {}^{-2}\right\rangle _{M,F,J} =1\ , \label{mWIbr}%
\end{eqnarray}
where account has been taken of the fact that if
in some reference frame with a gauge Boson $F(\phi)$
the broken BRST-antiBRST symmetry term has
the form $M_F=M$ then in a different reference frame
determined by a gauge Boson $(F+\Delta F)(\phi)$
it should be calculated as
\begin{equation}\label{mfdf}
M_{F+\Delta F} = M_F + \Delta_{\lambda(\Delta F)} M_F\,.
\end{equation}
Here, the symbol \textquotedblleft$\langle\mathcal{O}\rangle_{M,F,J}%
$\textquotedblright\ for a quantity $\mathcal{O}=\mathcal{O}(\Gamma)$ stands
for a source-dependent average expectation value corresponding to a
gauge-fixing $F(\phi)$ with respect to $Z_{M,F}(J)$:
\begin{equation}
\left\langle \mathcal{O}\right\rangle _{M,F,J}=Z_{M,F}^{-1}(J)\int d\Gamma
\ \mathcal{O}\left(  \Gamma\right)  \exp\left\{  \frac{i}{\hbar}\left[
\mathcal{S}_{F}\left(  \Gamma\right)+M_F  +J_{A}\phi^{A}\right]  \right\}
\ ,\ \ \mathrm{with\ \ }\left\langle 1\right\rangle _{M,F,J}=1\ . \label{aexv1}%
\end{equation}
Owing to the presence of $\lambda_{a}(\Lambda
)$, and therefore also of $\Lambda(\Gamma)$,  the modified Ward identity depends on a choice of the gauge Boson
$F(\phi)$ for non-vanishing $J_{A}$. For constant $\lambda_{a}$, the identity (\ref{mWIbr})
decomposes in powers of $\lambda_{a}$ and assumes the form of two independent and one dependent
(at $\lambda^2$) identities:
\begin{eqnarray}
&& \left\langle \left[J_{A}\phi^{A}+M_F\right]  \overleftarrow
{U}^{a}\right\rangle _{M,F,J} =0\ , \label{mWIbr1}\\
&&  \left\langle   \left[J_{A}\phi^{A}+M_F\right]\overleftarrow{U}^{2}  -\left(  \frac{i}{\hbar}\right)  %
\left[J_{A}\phi^{A}+M_F\right]\overleftarrow{U}^{a}\left[J_{B}\phi^{B}+M_F\right]\overleftarrow{U}_{a}%
\right\rangle _{M,F,J} =0\,.
\label{mWIbr2}
\end{eqnarray}
The identities (\ref{mWIbr}), (\ref{mWIbr1}), (\ref{mWIbr2}) are different
from the Ward identities for a gauge theory without BRST-antiBRST symmetry breaking terms.

With account taken of (\ref{funcdeplafin}) and (\ref{mfdf}), the equality (\ref{mWIbr}) implies
a relation which describes the gauge dependence of $Z_{M,F}(J)$ for a finite
change of the gauge $F\rightarrow F+\Delta F$:%
\begin{align}
Z_{M_{F+\Delta F},F+\Delta F}(J) -Z_{M,F}(J) &  = Z_{M,F}(J)\left\langle \frac{i}{\hbar}%
J_{A}\phi^{A}\left[  \overleftarrow{U}^{a}\lambda_{a}\left(  \Gamma|-\Delta
{F}\right)  +\frac{1}{4}\overleftarrow{U}^{2}\lambda^{2}\left(  \Gamma
|-\Delta{F}\right)  \right]  \right.  \nonumber\\
&  -  \left.  (-1)^{\varepsilon_{B}}\left(  \frac{i}{2\hbar}\right)
^{2}J_{B}J_{A}\left(  \phi^{A}\overleftarrow{U}{}^{a}\right)  \left(  \phi
^{B}\overleftarrow{U}_{a}\right)  \lambda^{2}\left(  \Gamma|-\Delta{F}\right)
\right\rangle _{M,F,J}  . \label{GDInew2}%
\end{align}
From (\ref{GDInew}) it follows that upon the shell determined by $J_A=0$
a finite change of the generating functional of Green's functions
with a term of broken BRST-antiBRST symmetry does not depend
on the choice of the gauge condition with respect to a finite change
of the gauge, $F\to F+\Delta F$. The same statement takes place
for the physical $S$-matrix, due to the equivalence theorem \cite{equiv}.

Notice that the gauge independence of the functional $Z_{M,F}(J)$
can be achieved by considering the broken BRST-antiBRST symmetry
term $M$ as a composite field, $M \to L \cdot M$, with an
additional external bosonic source $L$ to $M$, however, achieved
in a way which takes into account the fact that $M$ is not the
same in different gauges, $F$ and $F+\Delta F$, which means that
$M_{F+\Delta F}= M_{F}+ \Delta M_{F}(\Delta F)$. The constancy of
$M$, $M_{F+\Delta F}= M_{F}$, at least for infinitesimal $\Delta
F$ has been used in \cite{lor} in a BRST-antiBRST formalism for $M
\equiv L_m\sigma^m(\phi)$ (with sources $L_m$ and composite fields
$\sigma^m$), which leads to the gauge independence of the
corresponding generating functional of Green's functions with
composite fields, $Z(J,\phi^*, \bar{\phi},L)$ in \cite{lor}, such
that $Z_{M,F}(J) = Z_{\sigma,F}(J,L) = Z(J,\phi^*,
\bar{\phi},L)\vert_{\phi^*=\bar{\phi}=0}$ only on the mass shell
determined in a larger space of sources (see (39), (40) therein)
$J_A=L_m=0$, with the subsidiary conditions $L_m=0$, as compared
to (\ref{GDInew2}). At the same time, if one takes into account
the non-constant character of the broken BRST-antiBRST symmetry
term $M$ in different gauges then, as has been demonstrated
\cite{Reshetnyak2} in Lagrangian BRST quantization, the gauge
independence of the generating functional of Green's functions
with composite fields (where the Gribov--Zwanziger horizon
functional $H$ is considered as a composite field, $H=\sigma$, see
Eqs. (18), (23) in \cite{Reshetnyak2}) is realized on the mass
shell $J=0$ determined only by the usual sources $J$ for arbitrary
$L$.

\section{Conclusion}

\label{Concl} In the present work, we have proved the correctness
(announced in \cite{MRnew2}) of the Jacobian (\ref{superJ1}) for a
change of variables in the partition function related to finite
field-dependent BRST-antiBRST transformations\footnote{As follows
from (\ref{eaBRSTantiBRST1}), (\ref{eaBRSTantiBRST2}), finite
BRST-antiBRST transformations with constant parameters form a
two-parametric Lie supergroup.} (\ref{Gamma_fin}) being polynomial
in powers of an Sp(2)-doublet of Grassmann-odd
functionally-dependent parameters $\lambda_{a}= \Lambda
\overleftarrow{U}_{a}$, generated by a finite even-valued
functional $\Lambda(\phi,\pi, \lambda)$ and nilpotent
Grassmann-odd operators $\overleftarrow{U}_{a}$ in (\ref{nilp}),
being the restriction of the generators $s_{a} $ of BRST-antiBRST
transformations  to the subspace $(\phi^{A},\pi
^{Ab},\lambda^{A})$. Based on the representation (\ref{superJ1}),
we justify the compensation equation for the parameters
$\lambda_{a}$ and find its solution (\ref{solcompeq2}),
(\ref{funcdeplafin}), which implies that for any finite change of
the gauge condition $F\to F+\Delta F$ there exists a unique
Sp(2)-doublet $\lambda_{a}(\Delta F)$ that ensures the gauge
independence of the vacuum functionals, $Z_F= Z_{F+\Delta F}$.
Conversely, any change of variables in the vacuum functional $Z_F$
given by a finite field-dependent BRST-antiBRST transformation
with functionally-dependent parameters $\lambda_{a}= \Lambda
\overleftarrow{U}_{a}$ is related to a finite change of the gauge
condition, $F\to F+\Delta F$, with $\Delta F(\Gamma|\Lambda)$
given by the solution (\ref{oppsol}) of the inverse problem for
the compensation equation (\ref{eqexp2}), so that the vacuum
functional $Z_F$ is gauge-independent. This follows from the fact
that the Jacobian of such a change of variables and the change of
the gauge are BRST-antiBRST-exact quantities. In the case of
functionally-independent odd-valued parameters $\lambda_a$, finite
BRST-antiBRST transformations generate in the corresponding
Jacobian some BRST-antiBRST-non-exact terms, which may not
correspond to a change of gauge-fixing in the quantum action. This
problem is studied in our forthcoming work \cite{MRmew4}. Note
that finite field-dependent BRST transformations do not imply such
a possibility. On the basis of the established representation for
the Jacobian (\ref{superJ1}), we justify the modified Ward
identity depending on the functionals $\lambda_{a}(\Delta F)$ in
(\ref{mWI}) and examine the problem of gauge-dependence for the
generating functional of Green's functions $Z_F(J)$ in
(\ref{GDInew}). As usual, albeit in the case of a finite change of
the gauge Boson $\Delta F(\Gamma|\Lambda)$, the functional
$Z_F(J)$ does not depend on a choice of the gauge on its the
extremals given by $J_A=0$, which is a basis for the study of
gauge-dependence for the effective action.

We have examined in more detail the properties of gauge theories with a closed algebra of rank 1,
including the Yang--Mills type of theories, on a basis of the above results
for a general gauge theory, and obtained a modified Ward identity
depending on the functional $\lambda_{a}(\Delta F)$ being restricted to the field
variables $\left(A^{i},B^{\alpha},C^{\alpha a}\right)$ and $(A^{m\mu},B^{m},C^{ma})$
in (\ref{mWIclalg}). In addition, we have described (\ref{GDInew1}) gauge-dependence
in these theories.

We have introduced the concept of BRST-antiBRST symmetry breaking in the
Sp(2)-covariant Lagrangian quantization and obtained the modified Ward
identity (\ref{mWIbr}) for the generating functional $Z_{M,F}(J)$ of Green's functions
with broken BRST-antiBRST symmetry. We have also studied gauge dependence
on the basis of finite field-dependent BRST-antiBRST transformations.
We have demonstrated that a non-renormalized functional $Z_{M,F}(J)$
does not depend on a choice of the gauge on the extremals given by $J_A=0$ in (\ref{GDInew2}),
thus providing (leaving aside the unitarity problem) a consistent
approach to introducing into a gauge theory of terms with broken BRST-antiBRST symmetry,
which has been applied in \cite{MRnew} to determine, in a consistent pertubative way,
the Gribov--Zwanziger horizon functional, obtained non-pertubatively in \cite{Zwanziger},
by using any other gauge beyond the Landau gauge, which may be compared
with other non-pertubative approaches \cite{sorellas,sorellas2}
to the Gribov--Zwanziger horizon functional in various gauges.

We have revised, as compared to the study of \cite{Reshetnyak,llr1,lrr}, the problem
of BRST symmetry breaking in the BV quantization and obtained the modified Ward
identity (\ref{mWIbvbr}) for the generating functional of Green's functions
with  broken BRST symmetry $Z_{M,\psi}(J)$ without introducing
a soft BRST symmetry condition for the BRST non-invariant term $M$.
We have also investigated gauge dependence by using the recently proposed
finite field-dependent BRST--BV transformations \cite{BLTfin}.
We have established that the non-renormalized functional $Z_{M,\psi}(J)$
does not depend on a choice of the gauge determined by
a Fermionic functional $\psi(\phi)$ on the extremals given by $J_A=0$
in (\ref{GDInewb}), thus justifying, in addition to \cite{Reshetnyak},
the consistency of introduction into a gauge theory
of terms with broken BRST symmetry in the framework of the BV quantization.

We have also introduced the concept of effective average action
within functional renormalization group approach in Yang--Mills
theories as an application of the general construction of
BRST-antiBRST symmetry breaking in the Sp(2)-covariant Lagrangian
quantization by specifying the broken BRST-antiBRST symmetry term.
To this end, we have introduced a regulator action (\ref{regSk})
and determined the generating functionals of Green's functions,
including the effective average action. For these functionals, we
have obtained the usual and modified (depending on the
field-dependent odd-valued parameters $\lambda_a$) Ward identities
with BRST-antiBRST symmetry breaking terms. We have established
that the non-renormalized functionals $\mathcal{Z}_{F;k}(J)$,
$\mathcal{Z}_{F;k}(J, \phi^*, \bar{\phi})$ do not depend on a
choice of the gauge determined by the functional $F(\phi)$ on the
extremals given by $J_A=0$ and on the hypersurface $\phi^*_{Aa}=0$
in (\ref{GDInew4}), thus providing a gauge-independent $S$-matrix
for any value of the external momentum-shell parameter, $k$. We
have proposed the form of the regulator action in any gauge
(\ref{lamaxi})--(\ref{mfdfymf}) from the $F_\xi$-family
(\ref{Fxi}) of gauges corresponding to the standard
$R_\xi$-gauges.

Notice that similar problems have been discussed independently
in \cite{BLTlfext}. As compared to our present work, the study of
\cite{BLTlfext} deals with a calculation of the Jacobian for a
change of variables given by BRST-antiBRST (BRST--BV by the
terminology of \cite{BLTlfext}) transformations with functionally-independent
field-dependent odd-valued parameters
$\lambda_a(\Gamma)$, subsequently used to formulate a compensation
equation, similar to (\ref{eqexpl}), but having a $2\times 2$
matrix form, which satisfies the condition of resolvability only
for functionally-dependent parameters, $\lambda_{a}=
\Lambda (\Gamma|\Delta F)\overleftarrow{U}_{a}$, whose form was first
announced in our work \cite{MRnew}.

In conclusion, note that among the interesting problems left
outside the scope of the present work, besides the evaluation of
Jacobians for arbitrary finite field-dependent BRST-antiBRST
transformations, one may turn to the study of the group properties
of finite field-dependent BRST-antiBRST transformations.

\section*{Acknowledgments}

A.A.R. is grateful to D. Bykov, D. Francia and to all the
participants of the International Seminar ``QUARKS 2014'' for
useful discussions. He also thanks the Organizing Committee for
the kind hospitality that made it possible to prepare the initial
part of this work. The study was supported by the RFBR grant under
Project No. 12-02-00121 and by the grant of Leading Scientific
Schools of the Russian Federation under Project No.  88.2014.2.
The work was also partially supported by the Ministry of Science
of the Russian Federation, Grant No. 2014/223.

\appendix

\section*{Appendix}

\section{On BRST Symmetry Breaking in Gauge Theories}

\label{AppA} \renewcommand{\theequation}{\Alph{section}.\arabic{equation}} \setcounter{equation}{0}

In this Appendix, we touch upon the study of \emph{BRST symmetry
breaking} in gauge theories, which is inspired by our works
\cite{Reshetnyak,llr1} within the BV quantization scheme \cite{BV},
and is based on the recently proposed BRST--BV field-dependent
transformations \cite{BLTfin}.
To this end, let us remind that the generating functional
of Green's functions $Z_{\psi}(J)$, implying the partition function
$Z_{\psi}= Z_{\psi}(0)$, is given by \cite{BV}
\begin{equation}\label{zpsi}
  Z_{\psi}(J) = \int d\phi\, d\phi^* d\lambda \;\exp\left\{
  \frac{i}{\hbar}
  \left[
{S}\left(  \phi, \phi^* \right) + \left( \phi^*_A - \psi(\phi)\frac{\overleftarrow{\delta}}{\delta \phi^A}  \right)\lambda^A + J_A\phi^A\right]  \right\},
\end{equation}
which, in terms of a nilpotent operator $\overleftarrow{U}$ introduced in \cite{rml1}, can be presented in the form
\begin{equation}\label{zpsi1}
  Z_{\psi}(J) = \int d\phi\, d\phi^* d\lambda \;\exp\left\{  \left(  i/\hbar\right)  \left[
{S}\left(  \phi, \phi^* \right) + \left( \phi^*_A\phi^A - \psi(\phi)  \right)\overleftarrow{U} + J_A\phi^A\right]  \right\}\ \ \mathrm{for} \ \
\overleftarrow{U} = \frac{\overleftarrow{\delta}}{\delta \phi^A} \lambda^A,
\end{equation}
where $\phi^A$, $\phi^*_A$, $\lambda^A$, $\psi(\phi)$ are the
respective fields, antifields, Lagrangian multipliers,\footnote{In
this Appendix, we use the standard notation \cite{BV} for
Lagrangian multipliers, which differs from the auxiliary fields
$\lambda^A$ of the opposite Grassmann parity in the basic part of
the work.} $\varepsilon(\phi^*_A)= \varepsilon(\lambda^A) =
\varepsilon_A+1$, and the gauge Fermion, which introduces the
gauge into the path integral. The quantum action $S$ is subject to
a master equation in terms of the antibracket $( \bullet , \bullet
)\equiv ( \bullet , \bullet )^1$ and the nilpotent Laplacian
$\Delta \equiv \Delta^1$ given by (\ref{abrack}) with $a=1$:
\begin{equation}\label{meq}
  \frac{1}{2}(S,S)=i\hbar\Delta S\  \Longleftrightarrow\
\Delta  \exp\left(  \frac{i}{\hbar}S\right)
=0 \ \mathrm{ for }\  S\vert_{\phi^*=\hbar=0}= S_{0}(A) %
\end{equation}
For vanishing external sources $J_A$, the integrand in
(\ref{zpsi}) or (\ref{zpsi1}) is invariant under the finite
BRST--BV transformations with a constant Fermionic parameter $\mu$
($\mu^2=0$),
\begin{equation}\label{fBRST}
  \Delta_\mu \left(\phi^A, \phi^*_A, \lambda^A \right) = \left(\phi^A \overleftarrow{U}, - \left(\phi^*_A , S\right), 0 \right)\mu \equiv \left(\phi^A, \phi^*_A, \lambda^A \right)
  \overleftarrow{s}\mu\,,
\end{equation}
whereas the transformation of the vacuum functional $Z_\psi$ with respect
to $(\phi^A, \phi^*_A, \lambda^A) \to  (\phi^A, \phi^*_A, \lambda^A)+ \Delta_\mu(\phi^A, \phi^*_A, \lambda^A)$,
related to finite field-dependent BRST-BV transformations \cite{BLTfin}
with an arbitrary functional $\mu=\mu(\phi,\lambda)$ is in one-to-one correspondence
with a finite change $\psi \to \psi +  \psi'$ of the gauge Fermion,
 \begin{equation}\label{psi'}
   \psi'(\phi,\lambda|\mu) = \frac{\hbar}{i}\left[ \sum
_{n=1}\frac{(-1)^{n-1}}{ n}\left( \mu\overleftarrow{U}\right)
{}^{n-1} \right]\mu\, ,
 \end{equation}
which justifies the gauge-independence of the vacuum functional, $Z_\psi = Z_{\psi+\psi'}$.
The relation (\ref{psi'}) is given by a solution of the so-called compensation equation
(implied by the condition $Z_\psi = Z_{\psi+\psi'}$) for an unknown functional
$\psi'$, which can be easily obtained from the Jacobian
$ J= \mathrm{Sdet}\left\Vert z^\mathrm{p}(1+\overleftarrow{s}\mu)
\frac{\overleftarrow{\delta}}{\delta z^{\mathrm{q}}}\right\Vert$,
for $z^{\mathrm{p}}\equiv (\phi^A, \phi^*_A, \lambda^A )$,
corresponding to the change of variables \cite{BLTfin}
\begin{equation}\label{compeqbv}
 \ln \left(1+ \mu \overleftarrow{U}\right)   = \frac{i}{\hbar}\psi' \overleftarrow{U}\ \  \mathrm{ with }\ \   J = \left(1+ \mu \overleftarrow{U}\right)^{-1}.
\end{equation}
Considering $\mu(\phi,\lambda)$ as an unknown parameter
of the field-dependent BRST--BV transformation (\ref{fBRST}),
which realizes $Z_\psi = Z_{\psi+\psi'}$ for a given finite change
of the gauge $\psi'$, a solution of (\ref{compeqbv}) with accuracy
up to $\overleftarrow{U}$-exact terms reads
\begin{equation}
\mu(\phi,\lambda|\psi')= - \frac{i}{\hbar}g(y)\psi'\ ,\ \ \mathrm{for}%
\ \ \ y\equiv\frac
{i}{\hbar}\psi'\overleftarrow{U}\ , \label{solcompeqBV}%
\end{equation}
with a function $g(y)$ given by (\ref{solcompeq2}).

Let us now consider a BRST non-invariant Bosonic functional
$M_\psi = M_\psi(\phi, \phi^*)$, $M_\psi \overleftarrow{s}\ne 0$,
with $\mathrm{gh}(M_\psi )=0$, which may, or may not, be subject
to the so-called condition of soft BRST symmetry breaking \cite{llr1,lrr}
\begin{equation}\label{brstbre}
  \left(M_\psi,M_\psi\right)= 0,\    \left[\left(M_\psi,M_\psi\right)= -2i\hbar\Delta M_\psi\right] , %
\end{equation}
and is assumed to be such that the path integrals $Z_{M_\psi,
\psi}(J)$ and $Z_{M_\psi, \psi}=Z_{M_\psi, \psi}(J)|_{J=0}$,
\begin{equation}\label{zpsibr}
Z_{M_\psi,\psi}(J) = \int d\phi\, d\phi^* d\lambda \;\exp\left\{  \left(  i/\hbar\right)  \left[
S_{M}\left(  \phi, \phi^* \right) + \left( \phi^*_A\phi^A - \psi(\phi)  \right)\overleftarrow{U} + J_A\phi^A\right]  \right\},\  \mathrm{for} \ Z_{0,\psi}(J)= Z_{\psi}(J),
\end{equation}
with the action
\begin{equation}\label{nbaction}
S_{M} = S+ M_\psi \,,\,\,\, \mathrm{so}\ \mathrm{that}\   S_{M=0}=S,
\end{equation}
be well-defined in perturbation theory.
In this case, we will speak
of a \emph{gauge theory with BRST symmetry breaking}
and refer to $Z_{M_\psi,\psi}(J)$ in (\ref{zpsibr}) as the generating functional
of Green's functions with BRST symmetry breaking.
It is obvious that the integrand in (\ref{zpsibr}) for $J=0$ is not BRST invariant.
However, the gauge independence of the vacuum functional
$Z_{M_\psi,\psi}(0)$ can be restored if we suppose that
within the reference frame given by the gauge Fermion $\psi+\psi'$
the BRST symmetry breaking term should have a BRST transformed
representation, with $\mu(\psi') = \mu(\phi,\lambda|\psi')$
being a solution (\ref{solcompeqBV}) of the compensation equation (\ref{compeqbv}), namely,
\begin{equation}\label{mpsidf}
M_{\psi+\psi'} = M_\psi + \Delta_{\mu(\psi')} M_\psi\,.
\end{equation}
Indeed, a modified Ward identity for $Z_{M_\psi,\psi}(J)$
is easily obtained by making in (\ref{zpsibr}) a field-dependent BRST transformation
(\ref{fBRST}) and using the relations (\ref{solcompeqBV}) and the expression
(\ref{compeqbv}) for the Jacobian:
\begin{equation}
\left\langle \left\{  1+\frac{i}{\hbar}\left[J_{A}\phi^{A} + M_\psi\right]\overleftarrow
{U}\mu(\psi')  \right\}  \left(  1+\mu(\psi')\overleftarrow
{U}\right)  {}^{-1}\right\rangle _{M,\psi,J} =1\,, \label{mWIbvbr}%
\end{equation}
where the symbol \textquotedblleft$\langle\mathcal{O}\rangle_{M,\psi,J}%
$\textquotedblright\ for a quantity $\mathcal{O}$ stands for a
source-dependent average expectation value with respect to
$Z_{M,\psi}(J)$, corresponding to the gauge-fixing $\psi$. Notice
that (\ref{mWIbvbr}) differs  from the Ward identity for a gauge
theory without BRST symmetry breaking and takes the form, with a
constant $\mu$,
\begin{eqnarray}
&& \left\langle \left[J_{A}\phi^{A}+M_{\psi}\right]  \overleftarrow
{U}\right\rangle _{M,\psi,J} =0\ , \label{mWIbvbr1}
\end{eqnarray}
which is identical, after introducing external antifields $\phi^*_A$,
to the Ward identity for $Z_{M_\psi,\psi}(J,\phi^*)$ in \cite{Reshetnyak,llr1}.

The Ward identity (\ref{mWIbvbr}), with allowance made for (\ref{solcompeqBV}) and (\ref{mpsidf}),
implies an equation which describes the gauge dependence of $Z_{M,\psi}(J)$ for a finite
change of the gauge $\psi\rightarrow \psi+\psi'$, namely,%
\begin{align}
Z_{M_{\psi+\psi'},\psi+\psi'}(J) -Z_{M,\psi}(J) &  = Z_{M,\psi}(J)\left\langle \frac{i}{\hbar}%
J_{A}\phi^{A} \overleftarrow{U}\mu\left(  \phi,\lambda|-\psi'\right)
\right\rangle _{M,\psi,J}  . \label{GDInewb}%
\end{align}
As in the case of BRST-antiBRST symmetry (\ref{GDInew}), the
relation (\ref{GDInewb}) allows one to state, due to the standard
arguments of the equivalence theorem \cite{equiv}, that upon the
shell determined by $J_A=0$ a finite change of the generating
functional of Green's functions with a broken BRST symmetry term
does not depend on a choice of the gauge condition with respect to
a finite change of the gauge $(\psi\to \psi+\psi')$. Note, first
of all, that this result is in complete agreement with the result
of \cite{Reshetnyak} when the total configuration space is
determined only by the fields $\phi^A$ without introducing the
internal antifields $\phi^*_A$ and Lagrangian multipliers
$\lambda^A$. Second, the same statement can be made for the
physical $S$-matrix. We hope that in the case of a renormalized
theory the above property is preserved by a renormalized
functional $Z_{M,\psi;R}(J)$ and intend to study this problem in a
forthcoming work.

\section{Effective Average Action in Yang--Mills theories}

\label{AppB}  \setcounter{equation}{0}

Here, we apply the concept of BRST-antiBRST symmetry breaking
to the effective average action in Yang--Mills  theories
within the functional renormalization group approach
(for a review and references, see \cite{Gies})
to the BRST-antiBRST Lagrangian quantization of Yang--Mills theories.

In this case, the generating functional (\ref{z(0)}),  with $S$ being
a solution of (\ref{3.3}), is reduced by integration over
$(\phi^*_a, \bar{\phi},\pi^a,\lambda)$ to
$Z_{F}(J)$ in (\ref{zj}), with the BRST-antiBRST quantum action
$S_{F}\left(  \phi\right)$ in (\ref{action}) invariant
under the finite BRST-antiBRST transformations
$\phi^{A} \to \phi^{A}+\Delta_\lambda\phi^{A}$ in (\ref{finite}).
We consider the  generating functional of Green's functions
extended by external antifields $\widetilde{\phi}^*_a, \widetilde{\bar{\phi}}$
and written without the tilde symbol as
$Z_{F}(J,\widetilde{\phi}^*, \widetilde{\bar{\phi}}) = Z_{F}(J,\phi^*, \bar{\phi})$,
\begin{align}
Z_{F}(J, \phi^*, \bar{\phi})  &  =  \int d\phi\ \exp\left\{  \frac{i}{\hbar}\left[
\mathcal{S}_{F}(\phi,\phi^*, \bar{\phi})   +J_{A}\phi^{A}\right]  \right\} ,\label{zje} \ \ \mathrm{with } \ \ \mathcal{S}_{F} = S_{F} + \phi^*_{Aa} s^a \phi^A - \frac{1}{2}  \bar{\phi}_A s^2 \phi^A \, ,
\end{align}
for $Z_{F}(J,0, 0) = Z_{F}(J)$.
The generating functional
of vertex Green's functions
(effective action) $\Gamma_F(\phi, \phi^*, \bar{\phi})$ is determined
in the BRST-antiBRST quantization \cite{BLT1, BLT2} by
a Legendre transformation
of $(\hbar/i)\ln Z_{F}(J,\phi^*, \bar{\phi})$ with respect to the sources $J_A$:
\begin{equation}
\Gamma _{F}(\phi, \phi^*, \bar{\phi}) = \frac{\hbar}{i}\ln Z_{F}(J,\phi^*, \bar{\phi})- J_A\phi^A
\,,\quad \phi^{A}=\frac{\hbar\delta}{i\delta J_{A}}\ln Z_{F}(J,\phi^*, \bar{\phi})  \,, \label{eab}
\end{equation}%
with $J_A$ expressed in terms of the average fields $\phi^{A}$
from $J = - \delta \Gamma _{F}/\delta \phi$. Note that the
non-renormalized quantitiy $\Gamma_F=\Gamma_F(\phi, \phi^*,
\bar{\phi})$ satisfies the usual Ward identities in terms of the
extended antibracket:
\begin{equation}
\frac{1}{2}\big(\Gamma _{F}, \Gamma _{F}\big)^a
+ V^a \Gamma _{F}  = 0  \,, \label{WIeab}
\end{equation}%
rewritten, by means of (\ref{eab}), from the Ward identities for
$Z_{F}(J,\phi^*, \bar{\phi})$: $\big(J_A \delta / \delta \phi^*_A
- V^a\big) Z_{F}(J,\phi^*, \bar{\phi}) =0$. For the vanishing
$\phi^*_a, \bar{\phi}$, we also obtain the usual effective action
for the Yang--Mills theory in the BRST-antiBRST formalism: $\Gamma
_{F}(\phi, 0, 0) = \Gamma _{F}(\phi)$. The actions $\Gamma
_{F}(\phi, \phi^*, \bar{\phi})$ and $\Gamma _{F}(\phi)$ are
invariant under the respective finite BRST-antiBRST
transformations of the \emph{average fields}:
\begin{eqnarray}\label{eaBRSTantiBRST1}
  && \Delta_\lambda \big(\phi^A, \phi^*_{Aa}, \bar{\phi}_A\big) = \Big(\phi^{A}\left[\exp\big(\overleftarrow{s}{}^{a} \lambda_{a}\big)-1
\right], 0 , (-1)^{\varepsilon_A}\bar{\phi}_A \overleftarrow{V}{}^a \lambda_{a} \Big)\   \  \mathrm{for} \  (\,\bullet \overleftarrow{s}{}^{a})  = \left(\ \bullet ,  \Gamma _{F}(\phi, \phi^*, \bar{\phi})\right)^a
 \\
&&
 \mathrm{and} \ \  \phi^A  \to \check{\phi}^A =  (1+  \Delta_\lambda)\phi^{A}{}\big\vert{}_{\phi^*= \bar{\phi} =0} \Longleftrightarrow \phi^A  \to \check{\phi}^A = {\phi}^A \exp\big(\overleftarrow{s}{}^{a} \lambda_{a}\big) {}\big\vert{}_{\phi^*= \bar{\phi} =0}.\label{eaBRSTantiBRST2}
\end{eqnarray}
The finite BRST-antiBRST transformations
(\ref{eaBRSTantiBRST1}), (\ref{eaBRSTantiBRST2})
are nothing else than the \emph{Lie group transformations}
of a field configuration $\phi^A$, as compared to the usual infinitesimal
BRST-antiBRST transformations determined by the basic element
$\overleftarrow{s}{}^{a} \lambda_{a}$ of the Lie algebra.
The finite (group) transformations (\ref{eaBRSTantiBRST1}),
albeit with the initial (not average) fields $\phi^A$
respect the invariance of the integrand in $Z_{F}(J,\phi^*, \bar{\phi})$
for $J=0$ in (\ref{zje}), with account taken of the identity
$s^{a}s^{b}=\left(  1/2\right)  \varepsilon^{ab}s^{2} $, for $a,b=1,2$.

We now extend the functional renormalization group (FRG)
construction, earlier applied only in the Lagrangian BRST
quantization, to the case of BRST-antiBRST quantization. In doing
so, we introduce, instead of $\Gamma_F$, the so-called average
effective action $\Gamma _{F;k}$ with an external momentum-shell
parameter, $k$, smoothly related to $\Gamma_F$,
\begin{equation}\label{limgk}
\lim_{k\rightarrow 0}\Gamma _{F;k}(\phi, \phi^*, \bar{\phi})\ =
\ \Gamma_F(\phi, \phi^*, \bar{\phi}) ,
\end{equation}%
in such a way that the  action $\mathcal{S}_{F}(\phi,\phi^*,
\bar{\phi}) $ [for the vanishing antifields, being the action
$S_{F}(\phi)$] in Yang--Mills theories should be extended by
BRST-antiBRST symmetry breaking terms, $M_F(\phi)$, having the
form of a regulator action, $S_{k}= S_{F;k}$, being quadratic in
the fields $A^{i}=A^{m\mu}(x)$ and the Sp(2)-doublet of
ghost-antighost fields ${{C}^{\alpha a}}={{C}^{m a}}(x)$ for
$M(\phi) =S_{k}(\phi)$,
\begin{eqnarray}
S_{F;k}(\phi) &=&\frac{1}{2}A^{i}A^{j}(R_{F;k})_{ij}+ \frac{1}{2}\varepsilon_{ab}%
(R_{F;k,gh})_{\alpha \beta }C^{\beta b} {{C}^{\alpha a}}(-1)^{\varepsilon_\alpha} \nonumber   \\
&=&\int d^{D}x\Big\{\frac{1}{2}A^{m\mu}(x)(R_{F;k})_{\mu \nu
}^{mn}(x)A^{\nu n }(x)+\frac{1}{2}\varepsilon_{ab}{{C}^{m a}}(x)(R_{F;k,gh})^{mn}(x){{C}^{n b}}(x)\Big\}, \label{regSk}
\end{eqnarray}%
with the Lorentz indices $\mu, \nu = 0,1,...,D-1$ of the Minkowsky
space $R^{1,D-1}$, the metric
$\eta_{\mu\nu}=\mathrm{diag}(+,-,\ldots,-)$, and the indices
$m,n,l = 1,\ldots,N^2-1$ of the Lie algebra $su(N)$; see also
(\ref{xyYM}), (\ref{R(A)}) in Section~\ref{WIGD2}. In
(\ref{regSk}), we have specified the condensed notations, so that
the regulator quantities $(R_{F;k})$, $(R_{F:k,gh})$ in the
reference frame with a gauge Boson $F$ have no dependence on the
fields, as well as satisfy the properties
$(R_{F;k})_{ij}=(-1)^{\varepsilon _{i}\varepsilon
_{j}}(R_{F;k})_{ji}$,
$(R_{F;k,gh})_{\alpha\beta}=(-1)^{\varepsilon _{\alpha}\varepsilon
_{\beta}}(R_{F;k,gh})_{\beta\alpha}$ and vanish as the parameter
$k$ tends to zero.

By definition, the regulator
action $S_{F;k}$ is not BRST-antiBRST invariant:
\begin{eqnarray}\label{sababsk}
&& \hspace{-2em}  S_{F;k} \overleftarrow{s}{}^a  =  \int d^{D}x \left\{A^{m\mu}(R_{F;k})_{\mu \nu
}^{mn}D^{\nu nl }C^{l a} +\varepsilon_{bc}(R_{F;k,gh})^{mn}{{C}^{m b}}\Big(  \varepsilon^{ca}B^{n}-\frac{1}{2}f^{n o l}%
C^{lc}C^{oa}\Big) \right\}  \ne 0 , \\
&& \hspace{-2em} \frac{1}{4} S_{F;k}\overleftarrow{s}{}^2  =  \frac{1}{4}\int d^{D}x\left\{- A^{m\mu}(R_{F;k})_{\mu \nu
}^{mn}  \Big(2D^{\nu n l}%
B^{l}+f^{n l o}C^{oa}D^{\nu lp}C^{pb}\varepsilon_{ba}\Big) +   \varepsilon_{ab}D^{\mu m o }C^{o a} (R_{F;k})_{\mu \nu
}^{mn}D^{\nu nl }C^{l b} \right. \nonumber \\
&& \hspace{-0em} \left.   -2 \varepsilon_{ab}{{C}^{m a}}(R_{F;k,gh})^{mn}f^{nlo}\Big( B^{o}C^{lb}+\frac{1}{6}f^{ors}C^{sc}C^{rb}%
C^{ld}\varepsilon_{dc}\Big) - \varepsilon_{ab}\varepsilon_{cd} \Big(\varepsilon^{ac}B^{m}+\frac{1}{2}f^{mlo}C^{oc}C^{la}\Big)(R_{F;k,gh})^{mn}\right. \nonumber \\
&& \hspace{+1em} \left.  \qquad   \times \Big(\varepsilon^{bd}B^{n}+\frac{1}{2}f^{npr}C^{rd}C^{pb}\Big)   \right\}, \label{s2babsk} \\
&& \hspace{-0em} \Longrightarrow \ \  \Delta_\lambda S_{F;k} =  S_{F;k} \left[\exp\big(\overleftarrow{s}{}^{a} \lambda_{a}\big)-1 \right], \label{dbabsk}
\end{eqnarray}
where in deriving (\ref{sababsk}), (\ref{s2babsk}) we have used
the readily established Leibnitz-like properties of the generators
of BRST-antiBRST transformations, $s^{a}$ and $s^{2}$, acting on
the product of any functionals $A$, $B$ with definite Grassmann
parities:
 \begin{align}
& \left(  AB\right)  \overleftarrow{s}{}^{a} =\left(  A\overleftarrow{s}{}^{a}\right)  B\left(  -1\right)
^{\varepsilon_{B}}+A\left(  B\overleftarrow{s}{}^{a}\right),  \quad  \left(  AB\right)  \overleftarrow{s}{}^{2} =\left(  A \overleftarrow{s}{}^{2}\right)  B-2\left(  A \overleftarrow{s}_{a}\right)
\left(  B \overleftarrow{s}{}^{a}\right)  \left(  -1\right)  ^{\varepsilon_{B}}+A\left(
B \overleftarrow{s}{}^{2}\right)  \,. \label{s2(AB)}%
\end{align}
We define extended generating functionals $\mathcal{Z}
_{F;k}(J)$ of Green's functions coinciding for $k\to 0$ with the respective ${Z}%
_{F}(J,\phi ^{\ast }, \bar{\phi})$ and ${Z}%
_{F}(J)$ in (\ref{zje}), according to (\ref{z(0)m}),
\begin{align}
\mathcal{Z}_{F;k}(J, \phi^*, \bar{\phi})  &  =  \int d\phi\ \exp\left\{
\frac{i}{\hbar}\left[  \mathcal{S}_{F}(\phi,\phi^*, \bar{\phi}) + S_{F;k}  +J_{A}\phi^{A}\right]  \right\} .\label{zjek}
\end{align}
Before taking the limit $k\rightarrow 0$, the integrand in (\ref{zjek}) for $J=0$
is not BRST-antiBRST invariant due to (\ref{sababsk})--(\ref{dbabsk}),
whereas in the limit $k\rightarrow 0$ the functionals $\mathcal{Z}_{F;k}$
take values coinciding with the usual generating functionals
$Z_F$. The average  effective action $\Gamma
_{F;k}=\Gamma _{F;k}(\phi ,\phi ^{\ast }, \bar{\phi} )$, being the generating functional
of vertex functions in the presence of regulators, is introduced
according to the rule described by (\ref{eab}), namely,
\begin{equation}
\Gamma _{F;k}(\phi ,\phi^{\ast }, \bar{\phi} )= \frac{\hbar}{i}\ln \mathcal{Z}_{F;k}(J,\phi^*, \bar{\phi})- J_A\phi^A
\,,\quad \phi^{A}=\frac{\hbar\delta}{i\delta J_{A}}\ln Z_{F;k}(J,\phi^*, \bar{\phi})  \,, \label{eabk}
\end{equation}%
with the obvious consequences
$J = - \delta \Gamma _{F;k}/\delta \phi$, for the Legendre transformation (\ref{eabk}).

The average effective action satisfies the functional integro-differential
equation
\begin{eqnarray}
\exp \Big\{\frac{i}{\hbar }\,\Gamma _{F;k}(\phi ,\phi^{\ast }, \bar{\phi} )\Big\} = \int
d\varphi \,\exp \Big\{\frac{i}{\hbar }\Big[\mathcal{S}_{F }({\phi +\hbar ^{\frac{1}{%
2}}\varphi },\phi^{\ast }, \bar{\phi})+S_{F;k}({\phi +\hbar ^{\frac{1}{%
2}}\varphi })\,  -\,\frac{\delta \Gamma _{F;k}(\phi ,\phi^{\ast }, \bar{\phi})}{\delta \phi }\,\hbar ^{%
\frac{1}{2}}\varphi \Big]\Big\}\,,  \label{GMEq-loop}
\end{eqnarray}%
determining the loop expansion $\Gamma _{F;k}=\sum_{n\geq 0}\hbar ^{n}\Gamma
_{n{}F;k}$. Thus, the tree-level (zero-loop) and one-loop approximations of (%
\ref{GMEq-loop}) correspond to
\begin{eqnarray}
&&\Gamma _{0{}F;k}(\phi ,\phi^{\ast }, \bar{\phi})\,=\,\mathcal{S}_{F {}0}(\phi ,\phi^{\ast }, \bar{\phi})\,+\,S_{0{}F;k}(\phi)\,,  \label{0loop} \\
&&\Gamma _{1{}F;k}(\phi ,\phi^{\ast }, \bar{\phi}) \,=\,\mathcal{S}_{F {}1}(\phi ,\phi^{\ast }, \bar{\phi})\,+ S_{1{}F;k}(\phi)-\frac{i}{2}\,\mbox{ln}\,\mathrm{Sdet%
}\left\Vert (\mathcal{S}_{F{}0}+\,S_{0{}F;k})^{^{\prime \prime }}_{AB}(\phi ,\phi^{\ast }, \bar{\phi})\right\Vert .  \label{1loop}
\end{eqnarray}
For the vanishing antifields $\phi^{\ast }_a, \bar{\phi}$,
(\ref{GMEq-loop})--(\ref{1loop}) imply the equation, as well as
the zero- and one-loop approximations, for the usual average
effective action $\Gamma _{F;k}(\phi) = \Gamma _{F;k}(\phi,
\phi^{\ast }, \bar{\phi})\vert_{\phi^{\ast } = \bar{\phi}=0}$ in
the BRST-antiBRST quantization.

Concerning the regulator functions, we suppose that they model the
non-perturbative contributions to the self-energy part of Feynman
diagrams, so that the dependence on the parameter $k$ enables one to extract
some additional information about the scale dependence of the theory beyond
the loop expansion \cite{Polch}.

The \emph{modified Ward identity}, or the \emph{Slavnov--Taylor identities},
for the functionals $\mathcal{Z}_{F;k}(J)$ and $\Gamma
_{F;k}(\phi)$ are readily obtained from the general result (\ref{mWIbr})
and have the form, for $ \mathcal{Z}_{F;k}(J)$,%
\begin{eqnarray}
&& \left\langle \left\{  1+\frac{i}{\hbar}\left[J_{A}\phi^{A}+S_{F;k}\right]\left[  \overleftarrow
{s}^{a}\lambda_{a}(\Lambda)+\frac{1}{4}\overleftarrow{s}^{2}\lambda
^{2}(\Lambda)\right]  -\frac{1}{4}\left(  \frac{i}{\hbar}\right)  {}^{2}%
\left[J_{A}\phi^{A}+S_{F;k}\right]\overleftarrow{s}^{a}\left[J_{B}\phi^{B}+S_{F;k}\right]\overleftarrow{s}_{a}%
\lambda^{2}(\Lambda)\right\}\right.\nonumber\\
&& \qquad \times\left.\left(  1-\frac{1}{2}\Lambda\overleftarrow
{s}^{2}\right)  {}^{-2}\right\rangle _{F;k,J} =1\ , \label{mWIbrk}%
\end{eqnarray}
with the source-dependent average expectation value corresponding to
the gauge-fixing $F(\phi)$ with respect to $\mathcal{Z}_{F;k}(J)$:
\begin{equation}
\left\langle \mathcal{O}\right\rangle _{F;k,J}=Z_{F;k}^{-1}(J)\int d \phi
\ \mathcal{O}\left(  \phi\right)  \exp\left\{  \frac{i}{\hbar}\left[
{S}_{F}\left( \phi \right)+S_{F;k}  +J_{A}\phi^{A}\right]  \right\}
\ ,\ \ \mathrm{for\ \ }\left\langle 1\right\rangle _{F;k,J}=1\ . \label{aexv3}%
\end{equation}
In (\ref{mWIbrk}), we have taken account of the fact that if in
some reference frame with a gauge Boson $F(\phi)$ the regulator
action has the form $S_{F;k}$ then in a different reference frame
determined by a gauge Boson $(F+\Delta F)(\phi)$ it should be
calculated in accordance with (\ref{mfdf}) and
(\ref{sababsk})--(\ref{dbabsk}) as follows:
\begin{equation}\label{mfdfym}
S_{F+\Delta F;k}
 = (1 + \Delta_{\lambda(\Delta F)}) S_{F;k} = S_{F;k}\exp\big( \overleftarrow{s}{}^a\lambda_a(\Delta F)\big)
\,.
\end{equation}
For   constant $\lambda_{a}$, the identity (\ref{mWIbrk})
decomposes in powers of $\lambda_{a}$ and assumes the form of two
independent and one dependent (at $\lambda^2$) identities
identical to (\ref{mWIbr1}), (\ref{mWIbr2}), where the
substitution $(M_F, \overleftarrow{U}^a, \langle ... \rangle
_{M,F,J} ) \to  (S_{F;k}, \overleftarrow{s}^a, \langle ... \rangle
_{F;k,J})$ has to be made.

For the average effective action $\Gamma_{F;k}(\phi)$,
the modified Ward identity (\ref{mWIbrk}) takes the form
\begin{eqnarray}
&& \left\langle\left\langle \left\{  1+\frac{i}{\hbar}\left[- \frac{\delta \Gamma _{F;k}}{\delta \phi^A} \phi^{A} + S_{F;k}\right]\left[  \overleftarrow
{s}^{a}\lambda_{a}(\Lambda)+\frac{1}{4}\overleftarrow{s}^{2}\lambda
^{2}(\Lambda)\right]  -\frac{1}{4}\left(  \frac{i}{\hbar}\right)  {}^{2}%
\left[- \frac{\delta \Gamma _{F;k}}{\delta \phi^A}\phi^{A}+S_{F;k}\right]\overleftarrow{s}^{a}\right.\right.\right.\nonumber\\
&& \qquad  \left.\left.\left.\times\left[- \frac{\delta \Gamma _{F;k}}{\delta \phi^B}\phi^{B}+S_{F;k}\right]\overleftarrow{s}_{a}%
\lambda^{2}(\Lambda)\right\} \left(  1-\frac{1}{2}\Lambda(\phi)\overleftarrow
{s}^{2}\right)  {}^{-2}\right\rangle\right\rangle _{F;k,\phi} =1\,, \label{mWIbrkg}%
\end{eqnarray}
where the operators  $\overleftarrow {s}^{a}$, $\overleftarrow
{s}^{2}$ do not act on $\delta \Gamma _{F;k}/\delta \phi$, and
$\langle\langle...\rangle\rangle_{F;k,\phi} $ denotes the average
expectation value corresponding to a gauge-fixing $F(\phi)$ with
respect to $\Gamma_{F;k}(\phi)$.

Repeating the arguments of Section~\ref{BaBsb} and using the
results for the Yang--Mills theory in the gauge dependence problem
(\ref{GDInew1}) of Section~\ref{WIGD2}, we obtain a relation which
describes the gauge dependence of $\mathcal{Z}_{F;k}(J)$ for a
finite change of the gauge $F\rightarrow F+\Delta F$, with account
taken of (\ref{funcdeplafin}) and (\ref{mfdfym}). Specified below
in (\ref{lDFYM}) for the Yang--Mills theory \cite{MRnew},  the
equality (\ref{mWIbrk}) implies a relation which describes the
gauge dependence of $\mathcal{Z}_{F;k}(J)$ with
a finite change of the gauge $F\rightarrow F+\Delta F$:%
\begin{align}
\mathcal{Z}_{F+\Delta F;k}(J) -\mathcal{Z}_{F;k}(J) &  = \mathcal{Z}_{F;k}(J)\left\langle \frac{i}{\hbar}%
J_{A}\phi^{A}\left[  \overleftarrow{s}^{a}\lambda_{a}\left(  \phi|-\Delta
{F}\right)  +\frac{1}{4}\overleftarrow{s}^{2}\lambda^{2}\left(  \phi
|-\Delta{F}\right)  \right]  \right.  \nonumber\\
&  -  \left.  (-1)^{\varepsilon_{B}}\left(  \frac{i}{2\hbar}\right)
^{2}J_{B}J_{A}\left(  \phi^{A}\overleftarrow{s}{}^{a}\right)  \left(  \phi
^{B}\overleftarrow{s}_{a}\right)  \lambda^{2}\left(  \phi|-\Delta{F}\right)
\right\rangle _{F;k,J}  , \label{GDInew3}%
\end{align}
where
\begin{equation}\label{lDFYM}
  \lambda_{a}\left(  \phi|-\Delta
{F}\right) = -\frac{1}{2i\hbar}\left[  \Delta F  \overleftarrow{s}_{a}
\right]  \sum_{n=0}^{\infty}\frac{(-1)^n}{\left(  n+1\right)  !}\left(  \frac
{1}{4i\hbar} \Delta F \overleftarrow{s}{}^2 \right)  ^{n}.
\end{equation}
From (\ref{GDInew3}) it follows that upon the shell determined by $J_A=0$
a finite change of the generating functional of Green's functions
with a  BRST-antiBRST-non-invariant regulator action $S_k$ does not depend
on a choice of the gauge condition with respect to a finite change
of the gauge, $F\to F+\Delta F$, for any value of the external
momentum-shell parameter $k$.

For the extended generating functional $\mathcal{Z}_{F;k}(J, \phi^*, \bar{\phi})$,
the finite (group) BRST-antiBRST transformations
$\phi^{A}\rightarrow\check{\phi}^{A} = \phi^{A}\exp\big( \overleftarrow{s}_{a}\lambda_a \big)$
in (\ref{eaBRSTantiBRST1}) within the sector of fields $\phi^A$ for an arbitrary functional
$\lambda_{a}(\phi)= \Lambda (\phi)\overleftarrow{s}_{a}$ result in
the modified Ward identity
\begin{eqnarray}
&&  \left\{ \left[J_{A}+\widehat{S}_{F;k}{}_A\right]\left[
\frac{\delta}{\delta\phi^*_{Aa}}\widehat{\lambda}_{a}(\Lambda)-\frac{1}{2}\frac{\delta}{\delta\bar{\phi}_{A}}\widehat{\lambda}
^{2}(\Lambda)\right] - V^a \widehat{\lambda}_{a}(\Lambda)  -\frac{\varepsilon_{ab}}{4}%
\left\{\left[\left[J_{A}+\widehat{S}_{F;k}{}_A\right]\frac{\delta}{\delta\phi^*_{Aa}} - V^a\right]\right.\right.\label{mWIbrke}\\
&& \left.\quad \times \left[\left[J_{B}+\widehat{S}_{F;k}{}_B\right]\frac{\delta}{\delta\phi^*_{Bb}}- V^b\right]%
\widehat{\lambda}^{2}(\Lambda)\right\} \left(  1-\frac{\varepsilon_{ab}}{2}\widehat{\Lambda},_{BA} \frac{\hbar\,\delta}{i\delta\phi^*_{Ab}}\frac{\hbar\,\delta}{i\delta\phi^*_{Ba}}(-1)^{\varepsilon_A+1}+ \widehat{\Lambda},_A \frac{\hbar\,\delta}{i\delta\bar{\phi}_{A}}\right)  {}^{-2}\mathcal{Z} _{F;k,J} =
0\ , \nonumber%
\end{eqnarray}
where the notation for $\widehat{S}_{F;k}{}_A$, $ \widehat{\lambda}_{a}$, $\widehat{\Lambda},_{A} $, $\widehat{\Lambda},_{BA} $ is
given by
\begin{eqnarray}
&&  \widehat{S}_{F;k},{}_A = \frac{\delta {S}_{F;k}(\phi)}{\delta\phi^A}\big|{}_{\phi \to \frac{\hbar}{i}\frac{\delta}{\delta J}},\ \  \widehat{\lambda}_{a} = {\lambda}_{a}(\phi)\big|{}_{\phi \to \frac{\hbar}{i}\frac{\delta}{\delta J}},\ \  \widehat{\Lambda},_{BA} =  \frac{\delta^2 \Lambda(\phi)}{\delta\phi^B\delta\phi^A}\big|{}_{\phi \to \frac{\hbar}{i}\frac{\delta}{\delta J}}. \label{widenot}%
\end{eqnarray}
For constant $\lambda_{a}$, the identity (\ref{mWIbrke}) decomposes
in powers of $\lambda_{a}$ and takes the form of two independent
(at $\lambda_a$) Ward identities:
\begin{equation}
  \widehat{q}^a \mathcal{Z} _{F;k,J} = 0\ \ \ \mathrm{for}\ \ \
\widehat{q}^a =  \left[J_{A}+\widehat{S}_{F;k}{}_A\right]\frac{\delta}{\delta\phi^*_{Aa}} - V^a , \label{mWIbr1e}
\end{equation}
with anticommuting $\widehat{q}^a $: $\widehat{q}^a\widehat{q}^b +
\widehat{q}^b\widehat{q}^a=0$, whereas at $\lambda^2$ the
corresponding Ward identity is implied automatically by
(\ref{mWIbr1e}), due to
\begin{equation}
 \frac{\varepsilon_{ab}}{4}%
\widehat{q}^a\widehat{q}^b =   -\frac{1}{2}\left[J_{A}+\widehat{S}_{F;k}{}_A\right]\frac{\delta}{\delta\bar{\phi}_{A}}  \,.
\label{mWIbr2e}
\end{equation}
For the extended average effective action $\Gamma_{F;k}(\phi,
\phi^*, \bar{\phi})$ in (\ref{eabk}), the \emph{modified Ward
identity} depending on the odd-valued functionally-dependent
functionals $\lambda_a = \Lambda \overleftarrow{s}_a$ can be
derived from (\ref{mWIbrke}) in a way similar to the trick in
\cite{llr1,Reshetnyak} within the BRST Lagrangian quantization;
however, in view of a non-linear character of the identity
(\ref{mWIbrke}) it is cumbersome and deserves a special analysis.
The identities (\ref{mWIbr1e}) for $\Gamma_{F;k}(\phi, \phi^*,
\bar{\phi})$ are readily obtained,
\begin{equation}
  \frac{1}{2}\big(\Gamma _{F;k}, \Gamma _{F;k}\big)^a
  + V^a \Gamma _{F;k} - {S}_{F;k}{}_A (\widehat{\phi}) \frac{\delta}{\delta\phi^*_{Aa}}
  \Gamma _{F;k} = 0\,, \label{mWIbr1ge}
\end{equation}
and coincide in the limit $k \to 0$ with (\ref{WIeab}) for the
usual effective action $\Gamma _{F}(\phi, \phi^*, \bar{\phi})$.
Here, the quantities ${S}_{F;k}{}_A (\widehat{\phi})$ for the
operator-valued fields $\widehat{\phi}{}^A$ are given by
\begin{eqnarray}
&&  {S}_{F;k},{}_A(\widehat{\phi}) = \frac{\delta {S}_{F;k}(\phi)}{\delta\phi^A}\big|{}_{\phi \to \widehat{\phi}}\,,
  \ \ \mathrm{with} \ \ \widehat{\phi}{}^A = \phi^A + i\hbar\,(\Gamma^{''-1}_{F;k})^{AB}
\frac{\delta_l}{\delta\phi^B}, \ \ \mathrm{for} \ \  (\Gamma^{''-1}_{F;k})^{AC}(\Gamma^{''}_{F;k})_{CB}=\delta^A_{\ B}
 , \label{widenot2}%
\end{eqnarray}
and $(\Gamma^{''}_{F;k})_{AB}\ =\ \frac{\delta_l}{\delta\phi^A}
\frac{\delta}{\delta\phi^B}\Gamma_{F;k}$.

Finally, a relation which describes the gauge dependence of the
extended functional $\mathcal{Z}_{F;k}(J, \phi^*, \bar{\phi})$ for
a finite change of the gauge $F\rightarrow F+\Delta F$  with
$\lambda_{a}\left(  \phi|-\Delta {F}\right)$ (\ref{lDFYM})
follows from (\ref{mWIbrke}) with account taken of (\ref{widenot}):%
\begin{align}
\mathcal{Z}_{F+\Delta F;k}(J,\phi^*, \bar{\phi}) -\mathcal{Z}_{F;k}(J\phi^*, \bar{\phi}) &  = \left\{ %
J_{A}\left[  \frac{\delta}{\delta\phi^*_{Aa}}\widehat{\lambda}_{a}\left(  -\Delta
{F}\right)  -\frac{1}{2} \frac{\delta}{\delta\bar{\phi}_{A}}\widehat{\lambda}{}^{2}\left(  -\Delta{F}\right)  \right] - V^a \widehat{\lambda}_{a}\left(  -\Delta
{F}\right) \right.  \nonumber\\
&  -  \left. \frac{\varepsilon_{ab}}{4}   \left[J_{A}\frac{\delta}{\delta\phi^*_{Aa}}  - V^a\right] \left[J_{B}\frac{\delta}{\delta\phi^*_{Bb}}  - V^b\right]\widehat{\lambda}^{2}\left(  -\Delta{F}\right)
\right\}\mathcal{Z}_{F;k}(J,\phi^*, \bar{\phi})  . \label{GDInew4}%
\end{align}
From (\ref{GDInew4}) it follows that upon the shell determined by $J_A=0$ on the hypersurface $\phi^*_{Aa} = 0$
a finite change of the extended generating functional
with a  BRST-antiBRST not-invariant regulator action $S_k$ does not depend
on a choice of the gauge condition with respect to a finite change
of the gauge, $F\to F+\Delta F$, for any value of the external
momentum-shell parameter $k$.

The consistency of the FRG method, based on the introduction of (\ref{sababsk}),
means that the values $\Gamma _{F;k}$, $\Gamma _{F+\Delta F;k}$
of the effective average actions
calculated in two different gauges determined by $\chi ^{a}$ and $\chi ^{a}+%
{\Delta }\chi ^{a}$ (see \cite{MRnew, BLT1} for details), corresponding  to
the gauge functionals $F $ and $F + {\Delta }F $, should
coincide with each other on the mass shell of vanishing antifields
$\phi^*_{Aa}$ for any value of the parameter $k$, i.e.,
along the FRG trajectory, but not only at its boundary points.
For completeness, the form of the FRG flow equation
for $\Gamma_{F;k}$ that describes the FRG trajectory
(formally identical with the one in \cite{LS}), with allowance
for the notation $\partial_t = k\frac{d}{dk}$, reads
\begin{equation}
\label{FRGeqbab}
\partial_t\Gamma_{F;k} \,=\, \partial_t S_{F;k}
\,-\, \frac{\hbar}{i}\,\,\Big\{
\frac{1}{2}\,\partial_t(R_{F;k})^{mn}_{\mu\nu}\,
\big(\Gamma^{''-1}_{F;k}\big)^{(m\mu)(n\nu)}
\,+\,\partial_t(R_{F;k,gh})^{mn}\,\big(\Gamma^{''-1}_{F;k}\big)^{mn}\Big\}\,
\end{equation}
and has the same form for the $\phi^*_{Aa}$- and $\bar{\phi}_A$-independent
part of $\Gamma_{F;k}$, due to the parametric dependence on
$\phi^*_{Aa}, \bar{\phi}_A$ for all the terms in (\ref{FRGeqbab}).

Let us consider a specific choice of the initial Bosonic gauge functional
$F = F_\xi$ from the $F_\xi$-family of gauge-fixing functionals $F_\xi=F_{
\xi  }\left(  A,C\right)  $, corresponding
to the $R_{\xi}$-like gauge introduced \cite{MRnew} in the
BRST-antiBRST Lagrangian quantization
\begin{eqnarray}
&& F_{  \xi  }    =\frac{1}{2}\int d^{D}x\ \left(  -A_{\mu}%
^{m}A^{m\mu}+\frac{\xi}{2}\varepsilon_{ab}C^{ma}C^{mb}\right)
\,,  \label{Fxi}\\
&& F_{ 0  }    = -\frac{1}{2}\int d^{D}x\ A_{\mu}^{m}A^{m\mu}%
\quad\mathrm{and}\quad F_{  1  }\ =\ \frac{1}{2}\int
d^{D}x\ \left(  -A_{\mu}^{m}A^{m\mu}+\frac{1}{2}\varepsilon_{ab}C^{ma}%
C^{mb}\right)  \ , \label{F01}%
\end{eqnarray}
where the gauge-fixing functional $F_{\left(  0\right)  }\left(  A\right)  $
induces the contribution $S_{F_0;k}$ to the quantum
action that arises in the case of the Landau gauge $\chi(A)=\partial^{\mu
}A_{\mu}^{m}=0$,  whereas the
functional $F_{ 1  }\left(  A,C\right)  $ corresponds to the
Feynman (covariant) gauge $\chi(A,B)=\partial^{\mu}A_{\mu}^{m}+\left(
1/2\right)  B^{m}=0$.
The parameters $\lambda_{a} (\phi|\Delta F(\xi))=s_{a}\Lambda$ of a finite
field-dependent BRST-antiBRST transformation that connects an $R_{\xi}$ gauge
with an $R_{\xi+\Delta\xi}$ gauge, according to (\ref{lDFYM}), have the form
\begin{eqnarray}
\!\!\!\lambda_{a}(\phi|\Delta F_\xi)  =\frac{\Delta\xi}{4i\hbar}\varepsilon_{ab}\int
d^{D}x\  B^{n}C^{nb}
\sum_{n=0}^{\infty}\frac{1}{\left(  n+1\right)  !}\left[
\frac
{\Delta\xi}{4i\hbar}\int d^{D}y\ \left(  B^{u}B^{u}-\frac{1}{24}%
\
f^{uwt}f^{trs}C^{sc}C^{rp}C^{wd}C^{uq}\varepsilon_{cd}\varepsilon
_{pq}\right)  \right]  ^{n}, \label{lamaxi}%
\end{eqnarray}
where allowance has been made for $\Delta F = \Delta F_\xi$,
\begin{equation}
\Delta F_{  \xi  }\ =\ F_{ \xi+\Delta\xi
}-F_{  \xi  }=\frac{\Delta\xi}{4}\varepsilon_{ab}\int
d^{D}x\ C^{ma}C^{mb},  \label{DFxi}%
\end{equation}
as well as for  $ \Delta F_{  \xi  }\overleftarrow{s}{}^a$
and $ \Delta F_{  \xi  }\overleftarrow{s}{}^2$,
\begin{eqnarray}
&& \hspace{-3em}\Delta F_{  \xi  }\overleftarrow{s}{}^a   =  \frac{\Delta\xi}%
{2}\varepsilon_{bc}\int d^{D}x\ \varepsilon^{ca}B^{m}C^{mb}  \ , \label{saDF} \\
&& \hspace{-3em}\Delta F_{  \xi  }\overleftarrow{s}{}^2  = 2 \int d^{D}x \left\{ \hspace{-0.1em} \left[ \hspace{-0.1em}  \left(  \partial^{\mu}A_{\mu}^{m}\right)
+\frac{\xi}{2}B^{m}\hspace{-0.1em} \right]\hspace{-0.1em}   B^{m}+\frac{1}{2}\left(  \partial^{\mu}%
C^{ma}\hspace{-0.1em} \right)  D_{\mu}^{mn}C^{nb}\varepsilon_{ab}-\frac{\xi}{48}%
\ f^{mnl}f^{lrs}C^{sa}C^{rc}C^{nb}C^{md}\varepsilon_{ab}\varepsilon
_{cd}\hspace{-0.1em}\right\}\hspace{-0.15em}  . \label{s2Fxi}
\end{eqnarray}
Now, with account taken of the representation (\ref{mfdfym}) for
the regulator action $S_{F_\xi;k}$ in the
$R_{\xi+\Delta\xi}$-gauge, one can find its form in a manner
respecting the gauge-independence problem described by
(\ref{GDInew3}), (\ref{GDInew4}) once $S_{F_\xi+\Delta F_\xi;k}$
is explicitly calculated by the rule
\begin{equation}\label{mfdfymf}
S_{F_\xi+\Delta F_\xi;k}
 =  S_{F_\xi;k}\exp\big[ \overleftarrow{s}{}^a\lambda_a(\Delta
 F_\xi)\big]
\,,
\end{equation}
with allowance for (\ref{sababsk})--(\ref{dbabsk}).

The result given by  (\ref{GDInew3}) and (\ref{GDInew4}) implies a
statement consistent with \cite{Reshetnyak} (as well as a
different statement in comparison with \cite{LS} for BRST
symmetry), i.e., on the gauge-dependence of the average effective
action, and therefore also on the consistency of its introduction
in Lagrangian quantization for any value of the parameter $k$.
Indeed, the explicit gauge dependence of the vacuum functional
$Z_{k,\chi}$ and of the average effective action  $\Gamma_{k}$ on
their extremals in \cite{LS} was shown respectively by Eqs. (4.12)
and (5.9) therein. At the same time, the gauge independence of the
average effective action in  \cite{LS} was achieved on the
mass-shell determined in a larger space of fields (see Eqs.
(6.31), (6.32) therein) with the subsidiary conditions $L=0$, by
consideration of the regulators $S_{k}$ as composite fields
($S_{k} \to L S_{k}$), however, without taking into account the
change of the regulators $S_{k}$ under the change of the gauge
condition in Eqs. (6.22), (6.27), (6.30), (6.31). In this respect,
note that the consideration of the regulators as composite fields,
following the approach \cite{Reshetnyak2} (see the remarks
 at the end
of Sect. \ref{BaBsb}), allows one to provide the gauge independence
of  $\Gamma_k$ on the mass shell given by the usual average
fields $\phi^A$ alone.

\end{document}